\documentclass[%
reprint,
groupedaddress,
frontmatterverbose,
showpacs,preprintnumbers,
amsmath,amssymb,
aps,
pra,
floatfix
]{revtex4-1}

\usepackage{graphicx,url,graphics}
\usepackage{dcolumn}
\usepackage{bm}
\usepackage{longtable}                                                                            
\begin{document}
\title{Valence-shell single photoionization of Chlorine-like K$^{2+}$ ions: Experiment and Theory}

\author{G.  A. Alna'Washi}\email{alnawashi@hu.edu.jo}
\altaffiliation{Present address: Department of Physics, The Hashemite University, Zarqa 13115, Jordan}
\author{M. Lu, M. Habibi},
\author{D. Esteves-Macaluso}\email{david.macaluso@umontana.edu}
\altaffiliation{Present address: Department of Physics and Astronomy, University of Montana, Missoula, Montana 59812, USA}
\author{J. C. Wang}
\author{R. A. Phaneuf}
\affiliation{Department of Physics, University of Nevada, Reno, NV 89557-0220}

\author{A. L. D. Kilcoyne}
\affiliation{Advanced Light Source, Lawrence Berkeley National Laboratory, 1 Cyclotron Road, Berkeley, CA 94720}

\author{C. Cisneros}
\affiliation{Instituto de Ciencias F\'{i}sicas, Universidad Nacional
		Aut\'onoma de M\'exico, Apartado Postal 48-3, Cuernavaca 62210, Morelos, M\'exico.}

\author{B. M. McLaughlin}\email{b.mclaughlin@qub.ac.uk}
\altaffiliation{Present address: Centre for Theoretical Atomic, Molecular and Optical Physics (CTAMOP), School of Mathematics and Physics,
                      The David Bates Building, 7 College Park, Queen's University of Belfast, Belfast BT7 1NN, United Kingdom}
\affiliation{Institute for Theoretical Atomic and Molecular Physics,
       		Harvard Smithsonian Center for Astrophysics,
		60 Garden Street, MS-14, Cambridge, MA 02138}
%
%

\date{\today}

\begin{abstract}
The absolute single photoionization cross-section was measured for Cl-like K$^{2+}$ over the photon energy
range from 44.2 - 69.7 eV at a constant energy resolution of 0.045 eV. The experiments were performed by
merging an ion beam with a beam of synchrotron radiation from an undulator.
The ground-state ionization threshold was measured at 0.004 eV energy
resolution to be 45.717 $\pm$ 0.030 eV. The measurements are rich in resonance structure
due to multiple Rydberg series of transitions to autoionizing states. These series are assigned spectroscopically
using the quantum defect method, guided by pseudo-relativistic Hartree-Fock
calculations for the energies and oscillator strengths of transitions to autoionizing states.
The experimental results, which include significant contributions from K$^{2+}$ ions
initially in metastable states, are in satisfactory agreement with a linear superposition
of semi-relativistic R-matrix calculations of photoionization cross sections from these initial states.
\end{abstract}



\maketitle
%
%
%
%

\section{Introduction}
Studies of photoionization of atomic ions lead to a fundamental understanding of atomic interactions occurring in the earth's
atmosphere \cite{Flower1983} and in high temperature environments such as stars, nebulae \cite{Hofmann1990} and controlled
thermonuclear fusion reactors \cite{McWhirter1983}. Experiments on photoionization of atomic and molecular ions have
become possible by utilizing the high photon flux of third-generation synchrotron radiation sources and the photon-ion merged-beams
technique \cite{West2001,West2004,Kjeldsen2006}, facilitating measurements at an unprecedented level of refinement and precision.
Photoionization cross sections of potassium ions have many applications in astrophysics.  Various lines of
K$^{2+}$ (K III) and K$^{3+}$ (K IV) ions have been detected in the planetary nebula NGC 7027
by the infrared Space Observatory Short Wavelength Spectrometer (SWS) \cite{Fuech1997}.  
An analysis using a simplified photoionization model with CLOUDY \cite{ferland1998,ferland2003} 
produced acceptable results when modeling the NGC 6302 neon line intensities from various ionization stages.
The model predicts a K VII  line to be the brightest of the K III to K VII series
of ionic lines. The measured upper limits for the lower-excitation [K] lines in the SWS spectrum of NGC 6302
are consistent with intensities predicted by the model.  However there is always room for improvement 
and  the need for more accurate atomic data is essential for these types of predictions. 
Furthermore, K III and K VI lines are also seen in the ultraviolet and visible spectra of the
symbiotic nova RR Telescopii \cite{Young2011,McKenna1997} and in the coronal line region of planetary nebulae
NGC 6302 and NGC 6537 \cite{Casassus2000}.

The photoionization process provides a highly selective probe of the internal electronic structure and dynamics of atoms,
molecules and their ions. Systematic studies along isoelectronic sequences are useful in predicting unknown
spectra for other members of the sequence. Strong electron-electron interactions introduce complexity to the
electronic structure of the chlorine isoelectronic sequence. Apart from the work of our group, only limited
experimental measurements have been carried out of photoionization cross sections for this isoelectronic sequence.
We note that preliminary studies of $2p$ photoabsorption in Cl, Ar$^+$, K$^{2+}$, and Ca$^{3+}$ ions were
made by Martins et al.  \cite{Fred2001} but full details were not published. This work was recently extended in a
detailed study of inner-shell photoionization of Cl by Stolte and co-workers \cite{Stolte2013}, who measured
relative partial ionization cross sections following photoexcitation of atomic chlorine near the Cl $2p$
and Cl $1s$ ionization thresholds.  Accompanying Breit-Pauli R-matrix theoretical calculations performed in the region
of the $2p$ thresholds showed suitable agreement with experiment.

Photoionization of atomic chlorine in the valence region was studied experimentally
by several groups using photoelectron spectroscopy \cite{Kimura1978, Berk1983,Krause1992, Krause1993}.
Alna'Washi and co-workers performed absolute cross-section measurements
for the Ca$^{3+}$ ion using the photon-ion merged-beams method that were in
satisfactory agreement with R-matrix calculations performed in intermediate coupling \cite{Phaneuf2010}.
Similarly, absolute photoionization cross-section measurements for Ar$^{+}$ ions by
Covington and co-workers  \cite{covington2001b,covington2011} using
the merged-beams technique were also in satisfactory agreement with theory. In that study,17 Rydberg series
due to $ 3p \rightarrow ns$ and $3p \rightarrow  nd$ converging to the $^1D_2$ and $^1S_0$ states of Ar$^{2+}$
were assigned~\cite{covington2011}.
We note that the $3p$ photoionization cross-section of Cl-like Potassium (K$^{2+}$)
has been calculated previously using the R-matrix theoretical approach in $LS$-coupling \cite{Taylor1978,rmat}, where the presence
of the $3s^23p^4(^1D_2)nd$ and $3s^23p^4(^1S_0)nd$ Rydberg series  were clearly illustrated.

Photoionization of atomic chlorine has been extensively studied theoretically during the last few
decades using a variety of approximations. These include R-matrix and K-matrix calculations carried out by
several groups \cite{Tayal1993, Felfi2002, Robicheaux1992,Armstrong1983}, the configuration-interaction method \cite{Martins2001},
many-body theory \cite{Samson1986}, open-shell transition-matrix \cite{Chang1984},
and an effective single-particle potential \cite{Kelly1986}.

This paper completes an investigation at the Advanced Light Source
of photoionization of ions of the Cl isoelectronic sequence. The absolute photoionization
cross-section was measured for K$^{2+}$ ions in the energy range 44.2 -- 69.7 eV.
Resonances observed in the photoionization cross section are assigned spectroscopically using quantum
defect theory (QDT) guided by pseudo relativistic Hartree-Fock calculations of energies
and oscillator strengths of autoionizing transitions (performed using the Cowan atomic structure code).
The measurements are compared directly with new R-matrix \cite{rmat}
theoretical results obtained in intermediate coupling using the Breit-Pauli approximation.
 %
%
%
%

\section{Experiment}
Absolute photoionization cross sections were measured using the merged-beams technique
on the ion-photon-beam (IPB) end station on undulator beam line 10.0.1.2 of the 
Advanced Light Source at Lawrence Berkeley National Laboratory.
A detailed description of the measurement technique was reported by Covington et al. \cite{covington2002}
and only a brief description is presented here. Atomic potassium was thermally evaporated into the discharge of
a 10-GHz permanent-magnet electron-cyclotron-resonance ion-source. Ions were extracted 
and accelerated by a potential difference of +6 kV, focused and collimated by a series of 
cylindrical einzel lenses and slits and magnetically analyzed according to their momentum-per-charge ratio.
A beam of $^{39}$K$^{2+}$ ions was selected, collimated and directed to a 90$^{\circ}$
electrostatic deflector, which merged it onto the axis of the highly collimated photon beam.
The latter was produced by an undulator and energy selected by a grazing-incidence spherical-grating monochromator.
A cylindrical einzel lens focused the ion beam at the center of the interaction region of length 29.4 cm. 
For absolute measurements, an electrical potential of +2 kV was applied to energy-label K$^{3+}$ 
product ions produced therein. Two-dimensional spatial profiles of the  merged ion and photon 
beams were measured by three translating-slit scanners at the beginning, middle and end points 
of the interaction region. Product K$^{3+}$ ions were separated from the primary K$^{2+}$ ion 
beam by a 45$^{\circ}$ demerger magnet. The primary beam was collected in an 
extended Faraday cup, while a spherical 90$^{\circ}$ electrostatic deflector directed the
 product ions onto a stainless-steel plate biased at -550 V, 
 from which secondary electrons were accelerated to a single-particle detector.
 The photoion yield was measured as the photon energy
 was stepped over the range 44.20 - 69.70 eV. Absolute photoionization cross-section measurements
 were performed at a number of discrete photon energies where
 no resonant features were present in the photoion-yield spectra.

 The absolute cross-section measurements were used to place the photoion-yield on an absolute cross-section scale.
 The total absolute uncertainty of the photoionization cross-section measurements
 in this experiment is estimated to be $\pm$ 20\%. The monochromator settings for the experiment with
 K$^{2+}$ ions was calibrated using the IPB end-station by re-measuring the $^2D^o_{3/2}$, $^2D^o_{5/2}$, $^2P^o_{1/2}$,
 and $^2P^o_{3/2}$ photoionization thresholds of Kr$^{3+}$, for which the energies
 48.79 $\pm$ 0.02 eV, 48.59 $\pm$ 0.01 eV, 46.91 $\pm$ 0.02 eV, and 46.62 $\pm$ 0.02 eV,
 respectively, were determined in a previous experiment \cite{Lu2006a}.
 The resulting uncertainty in the photon energy scale for the present K$^{2+}$ measurements is 
 conservatively estimated to be $\pm$~0.030 eV.
%
%
%
%
%
\section{Theory}
\subsection{R-matrix calculations}
 For comparison with the high-resolution measurements, state-of-the-art theoretical methods using
highly correlated wave functions with the inclusion of relativistic effects are required, since
fine-structure effects are resolved in the experiments. R-matrix \cite{rmat,Burke2011}
calculations of the photoionization cross-sections for the K$^{2+}$ ion 
were performed in intermediate coupling using an efficient parallel version of the R-matrix codes
\cite{Ballance2012} within the confines of a semi-relativistic Breit-Pauli approximation \cite{rmat,Burke2011}.
For the photoionization calculations on this system 30 $LS\Pi$ states (58 $J\Pi$ states) were included in the close-coupling
expansion arising from the following n=3 and 4 states of the
residual K$^{3+}$ ion core, namely, $1s^22s^22p^63s^23p^4 [^3P, ^1D, ^1S]$, $1s^22s^22p^63s3p^5 [^{1,3}P^o]$,
$1s^22s^22p^63s^23p^3(^4S^o, ^2D^o, ^2P^o)3d [^{1,3,5}L^o, L=0,1,2,3]$,
$1s^22s^22p^63s^23p^3(^4S^o, ^2D^o, ^2P^o)4s [^{1,3}P^o, ^{1,3}D^o, ^{3,5}S^o]$
and $1s^22s^22p^63p^6 [^1S]$. The orbital basis set employed for the residual K$^{3+}$ product ion was limited to n=4 in constructing
the multi-reference-configuration-interaction  wave functions used in our work. The Breit-Pauli approximation was used to calculate
 the energies of the 58 $J^{\pi}$ levels of the K$^{3+}$ residual ion arising from the above 30 $LS\Pi$
 states and all the subsequent K$^{2+}$ photoionization cross-sections.
 A minor shift (less than 0.5 \%) of the theoretical energy levels for the K$^{3+}$ residual ion was made in order to be in agreement with
 relativistic Hartree-Fock calculations \cite{Hansen1986} that are within 0.5 \% of the tabulated values \cite{NIST,Biemont1999,Sansonetti2008}.  
 
 Photoionization cross-sections were calculated for the $3s^23p^5 (^2P^o_{3/2})$
 ground state and all the following metastable states
 $3s^23p^5 (^2P^o_{1/2})$, $3s^23p^43d (^4D_{7/2,5/2,3/2})$,  $3s^23p^43d (^4F_{9/2,7/2,5/2})$,
 $3s^23p^43d (^4P_{5/2,3/2,1/2})$, $3s^23p^43d (^2F_{7/2,5/2})$,  $3s^23p^4 3d (^2G_{9/2,7/2})$ and
 $3s^23p^44s (^4P_{5/2,3/2,1/2})$ of the K$^{2+}$ ion in intermediate coupling.
\begin{table*}
\caption{\label{dipole}Dipole transitions and ionization potentials (eV) for  the ground and excited metastable states
               of K$^{2+}$ (K III) considered in the present R-matrix calculations.  Photoionization cross-section calculations were not
               carried out from the levels associated with the excited  $3s3p^6~ ^2S_{1/2}$,
              $3s^23p^4(^1D)3d^{\prime}~ ^2P_{3/2,1/2}$ and $3s^23p^4(^1D)3d^{\prime}~ ^2D_{5/2,3/2}$ states
              as they are dipole allowed to the $3s^23p^5~ ^2P^{\circ}_{3/2,1/2}$ lower levels.
              The ionization thresholds are given in eV for the K$^{2+}$ (K III) ion. NIST tabulated values
              are included for comparison purposes as are detailed atomic structure calculations using the GRASP code. 
              The percentage differences $\Delta_1$ (\%)  and $\Delta_2$   (\%) with
              the NIST values (where available) are respectively the Dirac and Breit-Pauli approximations.}
 \begin{ruledtabular}
\begin{tabular}{lcccccccccc}
\\
Configuration			&Term  		& GRASP$^{a}$&NIST 		& R-matrix$^{b}$ 	&$\Delta E^{c}$&$\Delta^e_1$	&$\Delta^f_2$	& $J^{\pi}$	& Label 			& $J^{\pi} \rightarrow J^{\prime \pi^{\prime}}$ transitions\\
					&			&(eV)		& (eV)		& (eV)	 		& (eV)		&(\%)	&(\%)	&			&				&								 \\
\\
\hline \\
 $3s^23p^5$ 			&$^2P^o$  	&45.8030		&45.8031		&44.9229	 		&-0.0001		&-0.0&-1.9	&${3/2}^o$ 	& --				&$3/2^o \rightarrow 1/2^e, 3/2^e, 5/2^e$ \\
					&			&45.8030		&45.717$^{\ddagger}$& 44.9229	&+0.0086		&-0.2&-1.8	&${3/2}^o$	&				&\\		
 					&			&44.4039$^{d}$& --			& --				& --			& --	& --		&${3/2}^o$	&				&\\		
 \\
 $3s^23p^5$ 			&$^2P^o$  	&45.5226		&45.5346		&44.6706			&-0.0120		&-0.7&-1.9	&${1/2}^o$ 	& --				&$1/2^o \rightarrow 1/2^e, 3/2^e$ \\
 \\
 $3s3p^6$ 			& $^2S$   		&28.3728		&29.6095		&28.2623			&-1.2367		&-4.4&-4.6	&${1/2}^e$	& First$^{\dagger}$	& \\
 \\
 $3s^23p^43d$ 		& $^4D$   		&25.6736		&--			&23.4033			&--			&--	&--		&${7/2}^e$	& First			&$7/2^e \rightarrow 5/2^o, 7/2^o, 9/2^o$ \\
 					&			&25.6475		&--      		&23.3838			&--			&--	&--		&${5/2}^e$ 	& First			&$5/2^e \rightarrow 3/2^o, 5/2^o, 7/2^o$ \\
					&			&25.6215		&--			&23.3603			&--			&--	&--		&${3/2}^e$	& First			& $3/2^e \rightarrow 1/2^o, 3/2^o, 5/2^o$ \\
 \\
 $3s^23p^43d^{\prime}$ & $^2P$		&23.5743		&22.8318		&23.2502			&+0.7425		&+3.2&+1.8	&${3/2}^e$	& Second$^{\dagger}$& \\
 		 			& 		  	&23.1555		&23.0051		&23.3444			&+0.1504		&+0.6&+1.5	&${1/2}^e$	& Second$^{\dagger}$ & \\
 \\
 $3s^23p^43d$			& $^4F$		&23.7927		& --      		&21.4413			&--			&--	&--		&${9/2}^e$	&First			&$9/2^e \rightarrow 11/2^o, 9/2^o, 7/2^o$ \\
                                  		&              		&23.6939		&--			&21,3556 			&--			&--	&--		&${7/2}^e$	&Second			 &$7/2^e \rightarrow 5/2^o, 7/2^o, 9/2^o$ \\
                                  		&              		&23.6222		&--			&21.2927 			&--			&--	&--		&${5/2}^e$	&Second			 &$5/2^e \rightarrow 3/2^o, 5/2^o, 7/2^o$\\
\\
 $3s^23p^43d$			&$^4P$	 	&22.5825		&--			&20.5340 			&--			&--	&--		&${5/2}^e$	&Third			&$5/2^e \rightarrow 3/2^o, 5/2^o, 7/2^o$ \\
                                 	 	&               		&22.6553		& --    		&21.2283			&--			&--	&--	         	&${3/2}^e$	 &Third			&$3/2^e \rightarrow 1/2^o, 3/2^o, 5/2^o$ \\
                                  		&               		&22.7135		& --    		&21.3949 			&--			&--	&--    		&${1/2}^e$	 &Third			&$1/2^e \rightarrow 1/2^o, 3/2^o$\\
 \\
 $3s^23p^43d^{\prime}$ & $^2D$		&22.1913		&21.9880		&20.1363			&+0.2033		&+0.9&-8.4	&${5/2}^e$	& Fourth$^{\dagger}$& \\
 		 			& 		   	&22.3457		&22.1324		&20.6083			&+0.2133		&+1.0&-6.9	&${3/2}^e$	& Fourth$^{\dagger}$ & \\
\\
 $3s^23p^43d$ 		&$^2F$		&22.0955		&-- 			&20.0613			&--			&--	&--		&${7/2}^e$	&Third			&$7/2^e \rightarrow 5/2^o, 7/2^o, 9/2^o$\\
                                 		&              		&21.8695		&20.8618		&19.8713 			&+1.0077		&+4.6&-4.8	&${5/2}^e$	& Fifth			 &$5/2^e \rightarrow 3/2^o, 5/2^o, 7/2^o$ \\
 \\
 $3s^23p^4 3d$		&$^2G$		&21.6998		&--			&19.3902			&--			&--	&--		&${9/2}^e$	& Second			& $9/2^e \rightarrow 11/2^o, 9/2^o, 7/2^o$\\
                                  		&              		&21.6867		&--			&19.3972 			&--			&--	&--		&${7/2}^e$	& Fourth			 &$7/2^e \rightarrow 5/2^o, 7/2^o, 9/2^o$\\
 \\
 $3s^23p^44s$ 		&$^4P$		&20.4312		& 20.0861		&18.3658			&+0.3451		&+1.7&-8.6	&${5/2}^e$	& Sixth			&$5/2^e \rightarrow 3/2^o, 5/2^o, 7/2^o$\\
                                  		&              		&20.2617		& 19.9291		&20.2696 			&+0.3326		&+1.6&+1.7        &${3/2}^e$	& Fifth			 &$3/2^e \rightarrow 1/2^o, 3/2^o, 5/2^o$\\
                                  		&             	 	&20.1595		& 19.8332		&20.6599 			&+0.3263		&+1.6&+4.2        &${1/2}^e$	& Fourth			 &$1/2^e \rightarrow 1/2^o, 3/2^o$\\
 \end{tabular}
\end{ruledtabular}
\begin{flushleft}
$^{a}$GRASP, $3s^23p^5$, $3s3p^6$, $3s^23p^43d$ and $3s^23p^4n\ell$ ($n$= 4, 5 with $\ell$ = $s$, $p$, $d$ and $f$) configurations used in the calculations.\\
$^{b}$Breit-Pauli R-matrix, closed channel bound - state calculations. \\ 
$^{c}$Energy difference (eV) of the GRASP values with the NIST tabulations.\\
$^{d}$MCDF ionization potential (eV) from Biemont and co-workers \cite{Biemont1999}.\\
$^{e}$Precentage difference of the GRASP values with the NIST tabulations.\\
$^{f}$Precentage difference of the Breit-Pauli values with the NIST tabulations.\\
$^{\dagger}$Allowed transition to the ground state not considered in the present excited state cross section calculations.\\
$^{\ddagger}$Ionization potential (eV) determined from the present experiment was found to be 45.717 $\pm$ 0.030 eV.
\end{flushleft}
\end{table*}
The cross-section calculations for photoionization from metastable states provide
insight into the initial-state distribution of the K$^{2+}$ primary ion beam.
Detailed structure calculations by Hibbert and co-workers \cite{Hibbert2000}
indicated these states lie between the  $3s^23p^5 (^2P^o_{3/2})$ ground state
threshold and the $3s^23p^44s (^4P_{5/2,3/2,1/2})$ excited states of the K$^{2+}$ ion.

The scattering wave functions were generated by allowing double-electron promotions out of the $n$=3 shell of the $3s^23p^5$
 base configuration into the orbital set employed.  Scattering calculations were performed with twenty continuum basis functions
 and a boundary radius of  ~14.537 Bohr radii. In the case of the $3s^23p^5  (^2P^o_{3/2})$ initial ground state, the dipole
 selection rule requires the dipole transition matrices, $3/2^o \rightarrow 1/2^e, 3/2^e, 5/2^e$, to be calculated.

 For  the ground and metastable states considered, the list of dipole matrices 
 for the various transitions are tabulated in Table \ref{dipole}
 together with their ionization potentials. The percentage difference compared 
 with the available NIST tabulated values is included to try and gauge the accuracy of the present Breit-Pauli results. 
  In Table \ref{dipole}  the calculated ionization potentials (using the closed-channel semi-relativistic 
  Breit-Pauli R-matrix approximation) is seen to differ from the NIST tabulated values by a few percent for 
  most of the levels and about 8\% for the higher lying excited states.
 
 We note in the absence of available NIST tabulated values (as is the case for many of the K III levels shown here) 
 the present Breit-Pauli results provides an estimate, particularly for the higher excited states. 
 As can be seen from Table \ref{dipole}, the finite basis set employed and the limited electron correlation included in the 
 collision model illustrates the difficulty of representing the energies of these excited states accurately. 
 This is evident from the higher lying excited metastable levels, with particularly the 
 $3s^23p^43d^{\prime}~^2D_{5/2}$ and $3s^23p^44s~^4P_{5/2}$ level splitting
deviating by approximately 8\% from the NIST tabulations.  It is not the remit of this 
paper to provide definitive values for the excited state ionization 
potentials within the present limited approximation.
Rather, the focus and aim of the present work is to try and 
provide an estimate and assess the contribution of resonance features 
in the photoionization cross sections from all the excited metastable levels considered here.
 Only extensive multi-configuration interaction  atomic structure calculations 
 (that includes fully relativistic effects), using progressively larger basis 
 sets and configuration interaction (CI) expansions
can address convergence to the definitive values for the ionization potentials 
and truly assess the accuracy of the present Bret-Pauli work.

Fully relativistic  structure calculations (shown in Table \ref{dipole}) for the 
K$^{2+}$ (K III)  ion were carried out using the Grant code 
GRASP \cite{dyall89,grant06,grant07}  with the $3s^23p^5$, $3s3p^6$, $3s^23p^43d$ 
 and $3s^23p^4n\ell$ ($n$= 4, 5 with $\ell$ = $s$, $p$, $d$ and $f$) configurations (205 levels),
We use these calculations to try  and assess the accuracy of 
 fine-structure excitation threshold levels obtained from the closed-channel Breit-Pauli R-matrix results.  
 As shown in Table \ref{dipole}, the fully relativistic structure calculations using the 
 GRASP code (with this larger basis set and CI expansion) for these same levels give much better agreement 
 with the tabulated NIST values.  The agreement is better than 4.6\% with many cases at the 2\% level.
 Extending the basis set  and CI expansions further would yield better agreement with the NIST values
 but be prohibitive to include in a collision model.
 The results from these relativistic structure calculations 
 performed with the GRASP code \cite{dyall89,grant06,grant07} shown in Table \ref{dipole} are seen 
 to provide a more accurate representation of the excited state metastable 
 threshold energies and the $j$-level splittings in the absence of experiment.   
 We note for the ground state ionization threshold the GRASP calculations are in excellent 
 agreement with the NIST value. 
   
 The Hamiltonian matrices for all the $1/2^o$, $3/2^o$, $5/2^o$, $7/2^o$, 
 $9/2^o$, $11/2^o$,  $9/2^e$, $7/2^e$, $5/2^e$, $3/2^e$, and $1/2^e$
 symmetries were then calculated, where the entire range of $LS$ matrices that contribute to these $J^{\pi}$ symmetries.
 For the initial $^2P^o_{3/2}$ ground state, the $^2P^o_{1/2}$ and all the metastable states listed in Table \ref{dipole},
  the electron-ion collision problem was solved  (in the resonance region below and between all the thresholds)
 using a suitably fine energy mesh of 5x10$^{-8}$ Rydbergs  (~0.68 $\mu$eV).
 These scattering calculations allowed complete resolution of the detailed resonance structure found in the appropriate
 photoionization cross sections for this ion.  Radiation damping was also included in our scattering calculations.
 The theoretical cross sections were convoluted with a Gaussian distribution having a profile of the
 same full-width at half maximum (FWHM) as that of experiment (45 meV) which
 enabled a direct comparison to be made with the
 experimental measurements.  
 To simulate the ALS experimental measurements, a non-statistical 
 averaging of the theoretical R-matrix photoionization cross-sections
was performed for the ground and the metastable states which
showed satisfactory agreement.
 
\subsection{Hartee-Fock calculations}
The Hartree-Fock approximation \cite{Friedrich1998} assumes that each electron in the atom
moves independently in the nuclear Coulomb field
and the average field of the other electrons and so the N-electron wave function 
is just the anti-symmetrized product of N one-electron spatial wave functions. 
As a guide in the assignment of resonant features in the measurements,
the Cowan atomic structure code \cite{Cowan1981}, which is based on the relativistic
Hartree-Fock (HFR) approximation, was used to calculate the energies and strengths of
excitations contributing to the photoionization cross-section for K$^{2+}$.
In the calculation of all transitions, $3s^23p^5$ was selected as the initial configuration.
The final configurations selected were $3s^23p^4ns$ (7 $\le n \le $ 20) for the $3p \rightarrow ns$ transitions,
$3s^23p^4nd$ (6 $\le n \le$ 20) for $3p \rightarrow nd$ transitions, and $3s3p^5np$ (4 $\le n \le$ 11) for $3s \rightarrow np$ transitions.
%
%
%
\section{Experimental Results and Analysis}

 \subsection{Overview of measurements}
\begin{figure}
\begin{center}
 \includegraphics[width=3.8in]{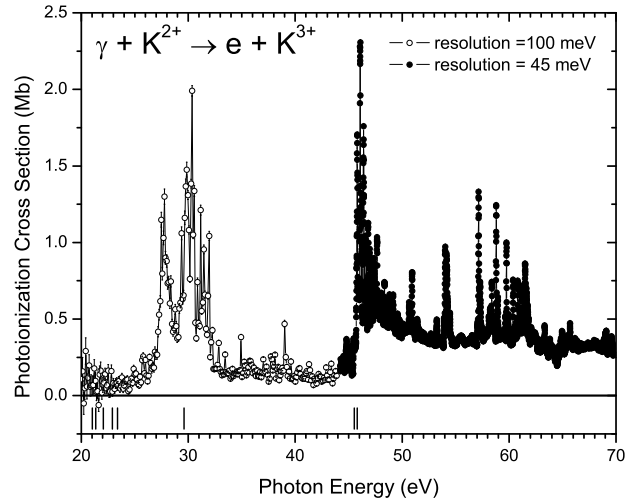}
 \caption{\label{overview} Overview of photoionization cross-section measurements over the energy range 20--70~eV. 
 					The data from 20--44~eV (open circles) have an energy resolution of 0.1~eV and 
					total uncertainty of $\pm$~50\%,  while those from  44--70~eV (solid black circles) 
					 have a resolution of 0.045~eV and uncertainty of $\pm$~22\%. 
 					The vertical lines indicate the ionization threshold energies listed in Table~\ref{dipole}. 
 					The highest in energy is for the ground state and the remainder are for metastable states.}
\end{center}
\end{figure}

 An overview of the photoionization cross-section measurements over the photon energy range from 20~eV to 70~eV 
 is presented in Fig.~\ref{overview}. The data from 20~eV to 44~eV were taken with an energy resolution of 0.1 eV, 
 and those from 44 eV to 70 eV with a resolution of 0.045~eV. Vertical lines in the figure indicate the ionization 
 threshold energies of the ground state (highest in energy) and different metastable states listed in Table~\ref{dipole}. 
 Evidently long-lived metastable states constituted a significant fraction of the primary K$^{2+}$ ion beam.

%

 \subsection{Ground-state ionization threshold}
\begin{figure}
\begin{center}
 \includegraphics[width=3.4in]{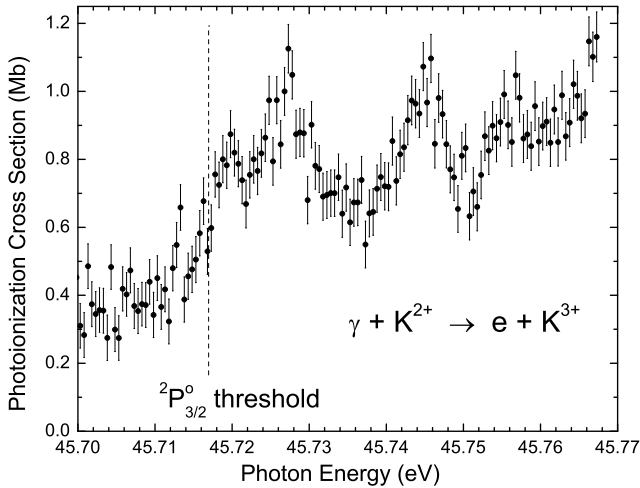}
 \caption{\label{3.8meV}  Photoionization cross-section measurements at 0.004 eV resolution
 			near the $^2P^o_{3/2}$ ground-state ionization threshold of K$^{2+}$ at 45.717 eV.}
\end{center}
\end{figure}

The ground-state configuration for the K$^{2+}$ ion is $1s^22s^22p^63s^23p^5$, for which the Russell-Saunders notation gives the
 terms $^2P^o_{3/2}$ for the ground state and $^2P^o_{1/2}$ for the metastable state. Figure~\ref{3.8meV} shows the photoionization cross section
 for K$^{2+}$ near the ground-state ionization threshold measured at a photon energy resolution of 0.004 eV. The ground-state
 ionization threshold was determined from the step to be 45.717 $\pm$ 0.030 eV. The measurement is 0.086~eV lower than the
 value tabulated in the NIST database \cite{NIST} of 45.803 $\pm$ 0.012~eV 
 which is the result from a multiconfiguration Dirac-Fock calculation~\cite{Biemont1999}.
 Using the tabulated value of 0.268 eV for the fine-structure splitting~\cite{NIST}
 gives 45.449 eV for the ionization threshold of the $^2P^o_{1/2}$ metastable state.
%
%

%
\begin{figure*}
\begin{center}
 \includegraphics[width=\textwidth, scale = 0.3]{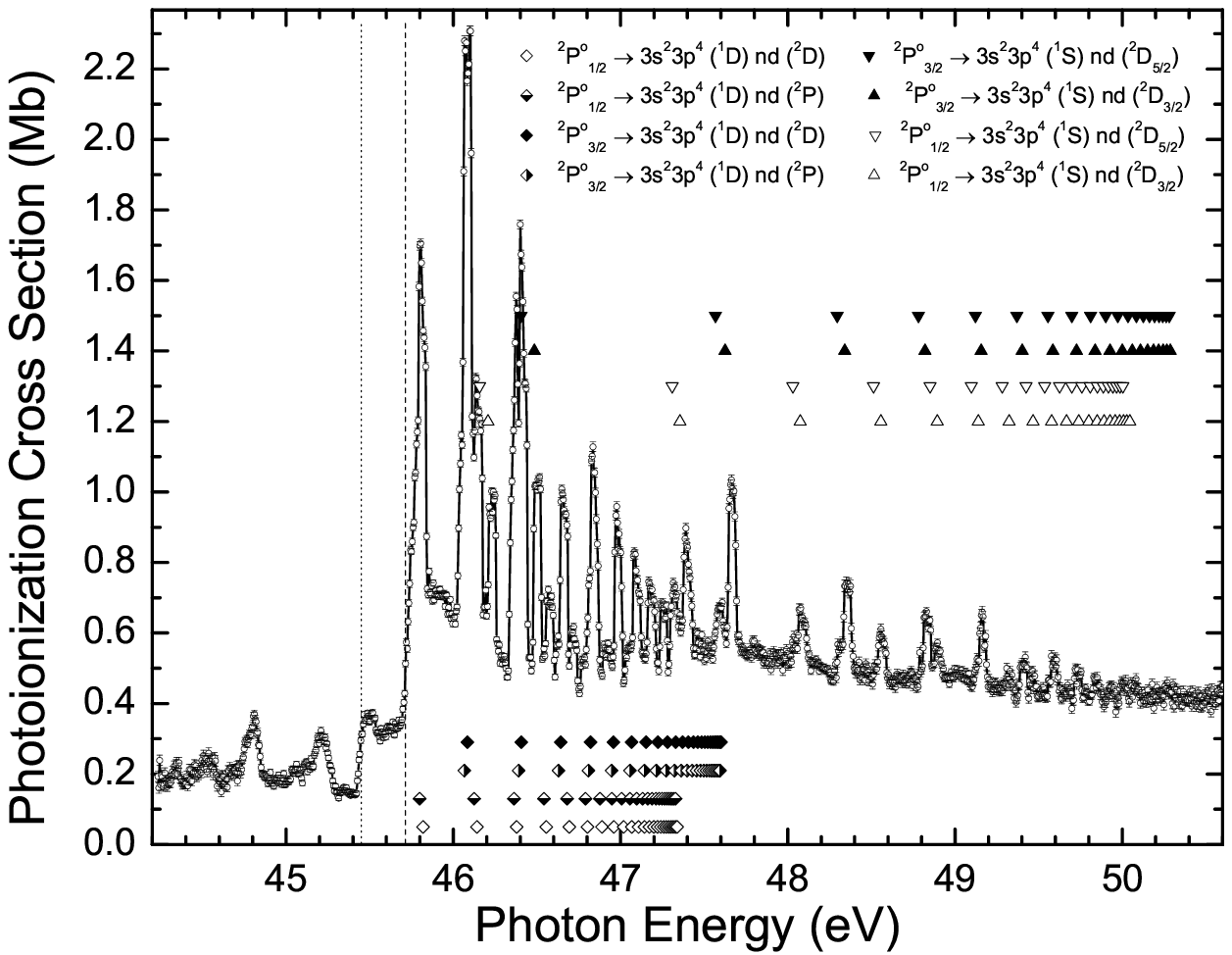}
 \caption{\label{45meVscan} Absolute cross-section measurements for  photoionization of K$^{2+}$ at a photon
 			energy resolution of 0.045 eV. Vertical dashed and dotted lines indicate
			the $^2P^o_{3/2}$ ground-state and $^2P^o_{1/2}$ metastable-state ionization
			threshold energies. Eight Rydberg series of resonances from the $^2P^o_{1/2}$
			metastable state and $^2P^o_{3/2}$ ground state of K$^{2+}$ converging to the
			$3s^23p^4(^1D_2)$ and $3s^23p^4(^1S_0)$ limits of K$^{3+}$ are identified.
			The measured cross section below the  $^2P^o_{1/2}$ threshold is attributed
			to population of higher lying quartet metastable states in the K$^{2+}$ ion beam.}
\end{center}
\end{figure*}
\subsection{$3p \rightarrow nd$ transitions}
Figure~\ref{45meVscan} shows the photoionization resonance structure in the energy range from  below the ionization threshold of the $^2P^o_{1/2}$ metastable
state to the $3s^23p^4(^1S)$ series limit of the K$^{3+}$ ion.  In this energy range eight Rydberg series due to $3p \rightarrow nd$ transitions differing in their
final coupling between the excited electron and the core are characterized and assigned spectroscopically using the quantum defect
form of the Rydberg formula. Of the eight series assigned in Figure~\ref{45meVscan}, four converge to the $3s^23p^4(^1D_2)$ limit of K$^{3+}$, two originating
from the $^2P^o_{1/2}$ metastable state and two from $^2P^o_{3/2}$ ground state. Transitions to the $3s^23p^4(^1D_2)nd(^2D^o)$ states (open diamonds)
and $3s^23p^4(^1D_2)nd(^2P^o)$ (half bottom filled diamonds) states from the $^2P^o_{1/2}$ metastable state are not fully resolved. Only the
lowest two members ($n$=9 and $n$=10) of the sequence are resolved from other Rydberg series. The corresponding series originating from the ground
state are also not resolved from each other, only the lowest member ($n$=9) is resolved from other Rydberg series.

The remaining four series converge to the $3s^23p^4(^1S_0)$ limit of K$^{3+}$. Two of these series originate from the $^2P^o_{1/2}$ metastable state
and two from $^2P^o_{3/2}$ ground state. In the two series $3s^23p^5(^2P^o_{1/2}) \rightarrow 3s^23p^4(^1S)nd(^2D^o_{3/2})$ (open triangles) and
$3s^23p^5(^2P^o_{1/2}) \rightarrow 3s^23p^4(^1S)nd(^2D^o_{5/2})$ (open inverted triangles), the first member ($n$=6) is not resolved from the other series,
while the next three members ($n$=7, 8, 9) are resolved because their energy position is above the $3s^23p^4(^1D_2)$ limit.
The situation is similar for the two corresponding series originating from the $3s^23p^5(^2P^o_{3/2})$ ground state.

The measured resonance energies, $E_n$ (eV)  of each of these series are plotted versus the principal quantum number
$n$,  in Figures~\ref{rydberg1} and \ref{rydberg2} in Appendix A and fitted to the quantum defect form of the Rydberg formula \cite{Seaton1983},
\begin{equation}
E_n = E_{\infty} - \frac{{\cal Z}^2Ry }{(n - \delta_n)^2}.
\end{equation}
Here, $n$ is the principal quantum number, $\delta_n$ the quantum defect,  being zero for a pure
 hydrogenic state.   The mean quantum defect is given by $\delta$ and the series limit $E_{\infty}$  (eV)
 are free parameters where ${\cal Z} = Z - N_c$. The Rydberg constant (Ry = 13.6057 eV), nuclear charge (Z = 19)
and the number of core electrons ($N_c$ = 16) are fixed parameters.
These eight Rydberg series are grouped together in Tables \ref{tab1} and \ref{tab2} in Appendix A by their energy positions,
quantum defects $\delta_n$, experimental series limits, and assignments.
The tabulated series limits in the NIST database \cite{NIST} for the $3s^23p^4(^1D_2)nd$ and $3s^23p^4(^1S_0)nd$ Rydberg series
are 47.834 eV and 50.582 eV, respectively. A comparison of these limits with the experimental limits in
Tables \ref{tab1} and \ref{tab2} provides additional evidence that the ground-state ionization threshold is 45.717 eV.

\subsection{Additional metastable states}
\begin{figure}
\begin{center}
\includegraphics[width=3.6in]{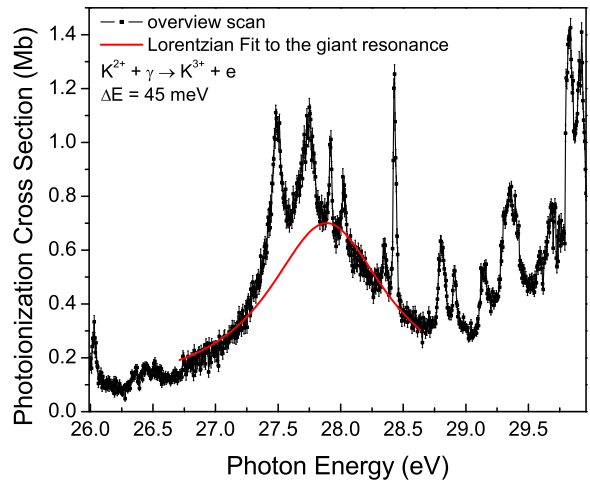}
 \caption{\label{overview26-30eV}(Color online) Absolute photoionization cross-section measurements for K$^{2+}$ at
					a photon energy resolution of 0.045 eV in the photon energy range 26 -- 30 eV.
 					The solid line represents a Lorentzian fit to the broad resonance feature attributed to 
					photoionization from the 3d$^4$P$_{5/2, 3/2, 1/2}$, 4s$^4$P$_{5/2, 3/2, 1/2}$ 
					and 3d$^4$G$_{9/2, 7/2}$ metastable states.}
\end{center}
\end{figure}

Below the ionization threshold of the $3s^23p^5(^2P^o_{1/2})$ metastable state at 45.450 eV,
Figure~\ref{45meVscan} shows a non-zero photoionization cross section and small resonance features,
suggesting population of more highly excited metastable states in the K$^{2+}$ ion beam.
Thus an overview energy scan at a photon energy resolution and energy step size of 0.1 eV was
made down to 20 eV, shown in Figure~\ref{overview}.  Strong resonance features are
evident above 24.4 eV. Therefore photoionization measurements in the energy
range 26 -- 30 eV were made at a photon energy resolution of 0.045 eV and step size of 0.005 eV, as
shown in Figure~\ref{overview26-30eV}. The spectrum is dominated by a
broad resonance feature of natural line width 1.15 eV, centered at 27.89 eV.
This broad feature was initially speculated to be due to  a  $3s^23p^4(^3P_{2,1,0})n\ell$ dipole resonance originating
from $3s^23p^4(^3P_{2,1,0})4s(^4P)$ metastable state. Narrow resonance features are superimposed
upon the broad resonance. The non-zero cross section measured below
the threshold of the $3s^23p^5(^2P^o_{1/2})$ metastable state is thus attributed
to the presence of an undetermined fraction of the primary ion
beam in the highly-excited $3s^23p^4(^3P)4s(^4P)$ metastable states,
which have an ionization threshold of 19.93 eV \cite{NIST}.

For an initial estimate the fraction of the  $3s^23p^4(^3P_{2,1,0})4s~^4P$
metastable states in the primary ion beam, the Cowan atomic structure code
was used to calculate the direct photoionization cross sections from these states.
The calculated direct photoionization cross section
from $3s^23p^4(^3P_{2,1,0})4s~^4P_{5/2,3/2,1/2}$, $3s^23p^5(^2P^o_{3/2})$
and $3s^23p^5 (^2P^o_{1/2})$ states near their ionization
thresholds are 0.14 Mb, 0.37 Mb, and 0.12 Mb.
Comparing these values to the measured non-resonant
photoionization cross sections, approximately 25\% of the primary K$^{2+}$ ion beam is estimated
to be in the $3s^23p^4(^3P_{2,1,0})4s~^4P$ metastable states, 25\% in the $3s^23p^5(^2P^o_{1/2})$
metastable state and 50\% in the $3s^23p^5(^2P^o_{3/2})$ ground state.
%
\begin{figure}
\begin{center}
 \includegraphics[width=3.4in]{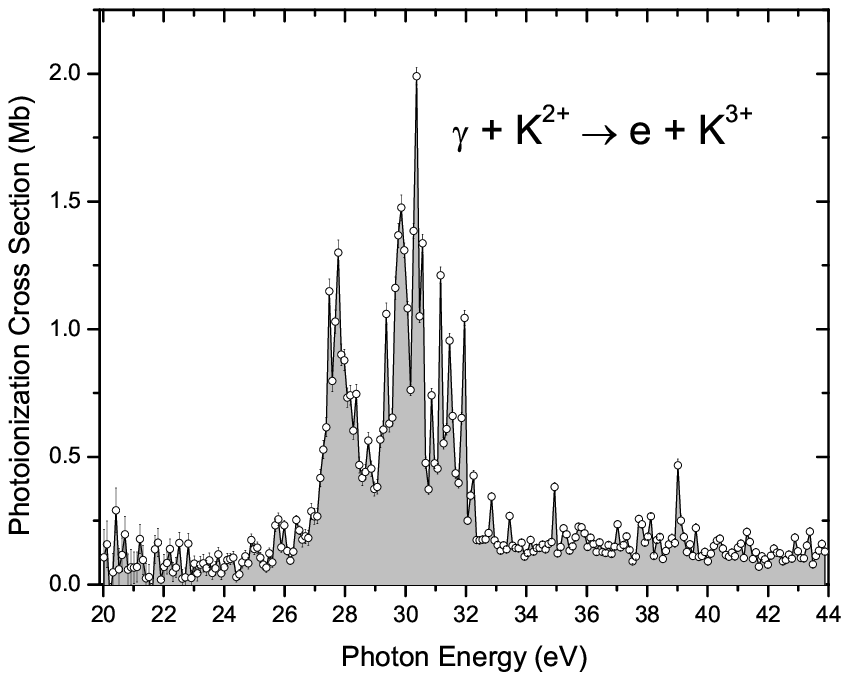}
 \caption{\label{overview2}Overview of photoionization cross-section measurements over the 
 				energy range 20 - 44 eV at a photon energy resolution of 0.1 eV and energy step size of 0.1 eV.}
 \end{center}
\end{figure}
%
%
%
%
\begin{figure*}
\begin{center}
 \includegraphics[width=\textwidth]{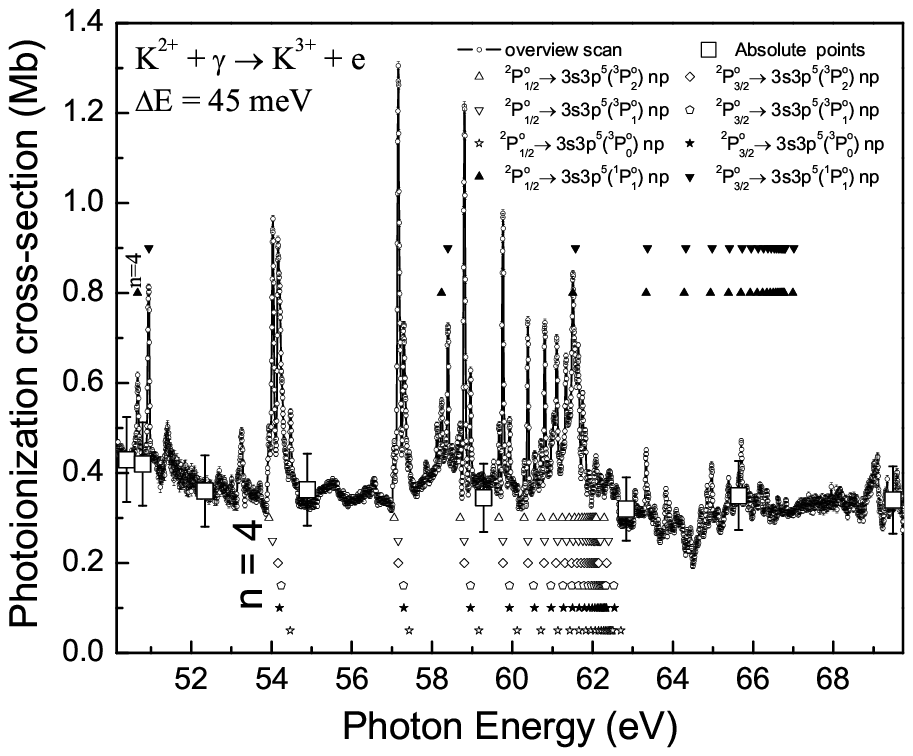}
 \caption{\label{8series} Absolute photoionization cross-section measurements for K$^{2+}$ at 0.045 eV
				       resolution in the energy range from 51.4 eV to 69.7 eV.
				       Open squares with error bars represent absolute measurements to which the energy scan is normalized.
 				       Eight Rydberg series of resonances due to inner-shell excitation
				       $3s \rightarrow np$ converging to the $^3P^o_2$, $^3P^o_1$, $^3P^o_0$,
				       and $^1P^o_1$ excited states of K$^{3+}$ are identified.}
\end{center}
\end{figure*}
%
\subsection{Inner-shell $3s \rightarrow np$ transitions}
Figure~\ref{8series} shows the photoionization resonance structure for K$^{2+}$ in the photon energy range 50.149 -- 69.741 eV where several series
of resonances due to $3s \rightarrow np$ inner-shell excitations are visible. Three Rydberg series are assigned to excitation of the
$3s3p^5(^3P^o_2)np$, $3s3p^5(^3P^o_1)np$ and $3s3p^5(^3P^o_0)np$ states from the $^2P^o_{1/2}$ metastable state, converging to the limits
of 62.291 $\pm$ 0.034 eV, 62.378 $\pm$ 0.034 eV, and 62.705 $\pm$ 0.067 eV, respectively. The three corresponding series originating
from the $^2P^o_{3/2}$ ground state converge to the limits of 62.356 $\pm$ 0.034 eV, 62.535 $\pm$ 0.034 eV, and 62.547 $\pm$ 0.034 eV, respectively.
A Rydberg series, $3s3p^5(^1P^o_1)np$, originating from the $^2P^o_{1/2}$ metastable state and converging to the series limit of
66.993 $\pm$ 0.049 eV, and a corresponding series originating from the  $^2P^o_{3/2}$ ground-state and converging to the series
limit of 67.017 $\pm$ 0.049 eV are also assigned in Figure~\ref{8series}.
The measured resonances of the eight Rydberg series in Figure~\ref{8series} are plotted versus the principal quantum number, $n$,
as shown in Figure~\ref{8series2} and fitted to the quantum defect form of the Rydberg formula \cite{Seaton1983} with mean quantum defect parameter $\delta$
and series limit $E_{\infty}$ as free parameters. These series are grouped in Tables \ref{tab3}, \ref{tab4}, and \ref{tab5}
by their measured energy positions, quantum defect parameters $\delta_n$, series limits, and assignments.
The tabulated series limits in the NIST database \cite{NIST} for
the series $3s3p^5(^3P^o_2)np$, $3s3p^5(^3P^o_1)np$, $3s3p^5(^3P^o_0)np$, and $3s3p^5(^1P^o_1)np$
are 62.440 eV, 62.623 eV, 62.721 eV, and 67.022 eV, respectively. A comparison of these limits with the experimental limits
in Tables \ref{tab3}, \ref{tab4}, and \ref{tab5} provides additional evidence that the ground-state ionization threshold is 45.717 eV.
An interesting question concerns the oscillator strengths of the assigned $3s3p^5(^3P^o_{2,1})4p$ and $3s3p^5(^3P^o_{2,1})5p$ resonances.
 It is assumed that for $n$ values higher than 4, the $^2P^o_{3/2} \rightarrow 3s3p^5(^3P^o_2)np$ resonances and
 $^2P^o_{1/2} \rightarrow 3s3p^5(^3P^o_1)np$ resonances are unresolved from each other.
Asymmetric Fano-Beutler resonance lineshapes \cite{Fano1968} are evident in Figure~\ref{8series} for the $3s \rightarrow np$ resonances.
This is attributed  to interference between the direct and the indirect photoionization
channels for excitation of the $3s$ subshell.

\section{Comparison with R-matrix Theory}


\begin{figure*}
\begin{center}
\includegraphics[width=\textwidth,scale=1.0]{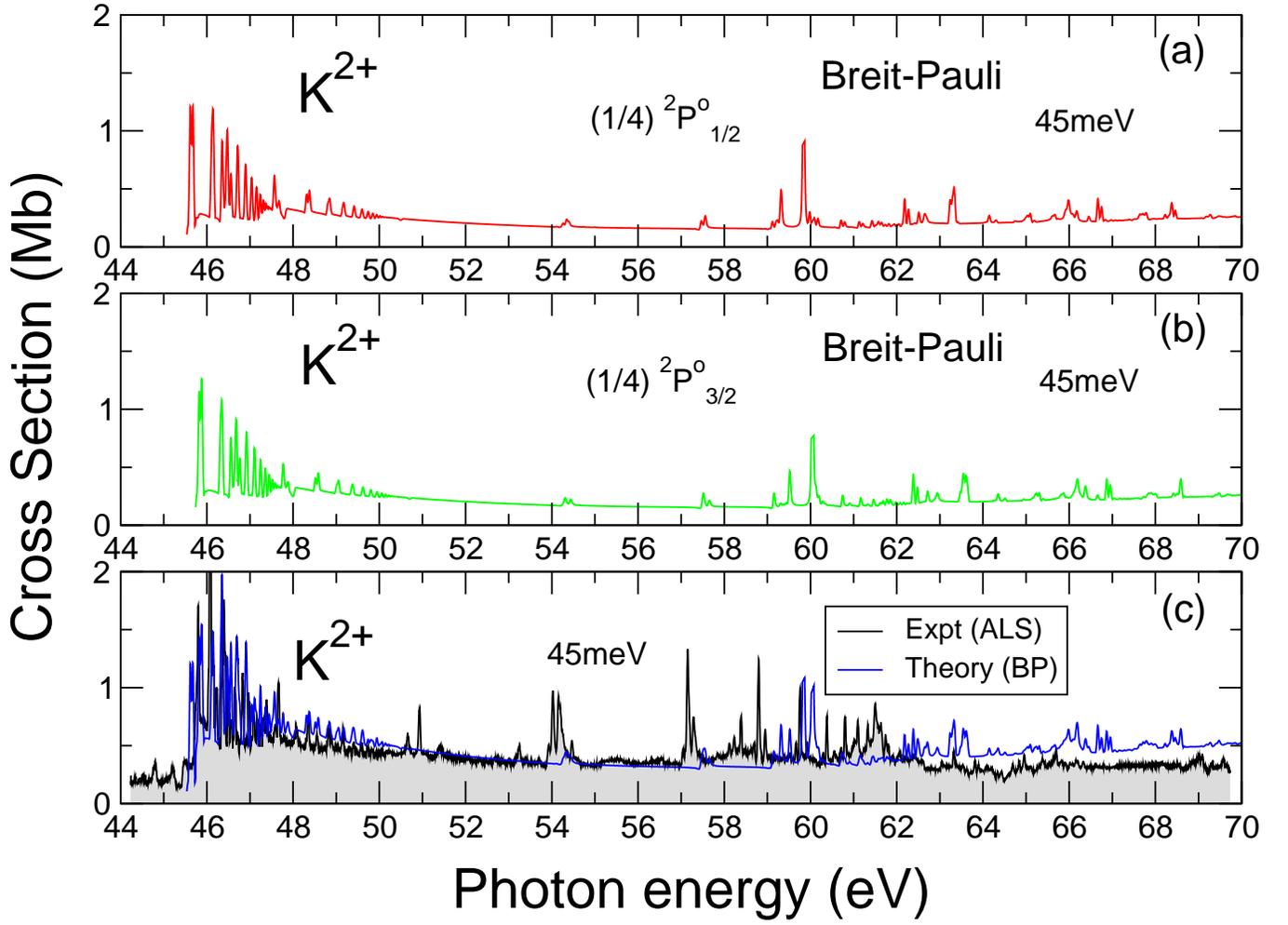}
\caption{\label{rmat17}   Breit-Pauli R-matrix photoionization cross section calculations
				  for the metastable  (a)(1/4) $3s^23p^5~^2P^{\circ}_{1/2}$
				   and (b) (1/4)$3s^23p^5~^2P^{\circ}_{3/2}$ ground state of K$^{2+}$
				   and (c) comparison between theory and experiment.
				   The theoretical results has been convoluted with a Gaussian distribution having a profile
				   of 45 meV FWHM to simulate the photon energy resolution of the experiment and a non-statistical
				   admixture of the ground state and metastable states in the primary K$^{2+}$ ion beam (see text for details).}
\end{center}
\end{figure*}
To further address the metastable content in the K$^{2+}$ parent ion beam,
semi-relativistic R-matrix calculations in intermediate coupling of cross sections
for photoionization were carried out from the ground state and all the
metastable states up to and including those from the $3s^23p^4(^3P)4s(^4P)$ levels.
Table \ref{dipole} lists all eighteen states investigated and the various excitation thresholds calculated
from the semi-relativistic R-matrix approach compared to the available experimental data.
Figures \ref{rmat11}  - \ref{rmat15} in Appendix B present the cross section as a
function of photon energy for these R-matrix calculations for the
ground and the metastable levels that are listed in Table \ref{dipole}.  For all of the metastable states investigated
up to and including those from the  $3s^23p^4(^3P)4s(^4P)$ levels, the R-matrix cross section calculations indicate
minimal presence of resonance features in the photon energy region 44 -- 70 eV.
However, resonances in the cross sections at photon energies below 44 eV are
 much stronger.   In the photon energy range 26 -- 30 eV
 one sees the main features are a broad shoulder resonance
 located at around 28 eV superimposed by peaks in the cross sections which are due
 to the presence of the $3d (^4P_{5/2,3/2,1/2})$,  $4s (^4P_{5/2,3/2,1/2})$ and
  $3d (^2G_{7/2,9/2})$ metastable states in the K$^{2+}$ primary ion beam.

A comparison in the photon energy range 44 -- 70 eV between the present experiment and the semi-relativistic
R-matrix intermediate-coupling cross-section calculations indicates best agreement with a non-statistical distribution among
the 18 possible initial states listed in Table \ref{dipole}.  Assuming 25\% population of the
$3s^23p^5(^2P^o_{3/2})$ ground  state and 25\% in the $3s^23p^5(^2P^o_{1/2})$
metastable state gives best agreement between theory and experiment, with the
remaining 50\% distributed among the more highly excited metastable states listed in Table \ref{dipole}.
Figure \ref{rmat17}  illustrates this comparison between the measurements
and semi-relativistic intermediate coupling R-matrix results.
The integrated oscillator strength $f$ from the R-matrix calculations is 0.106, which compares favorably with the
experimental value of 0.097 $\pm$ 0.021.

\section{Summary and Conclusions}
Absolute photoionization cross-section measurements for K$^{2+}$ were performed at fixed photon energy resolution of 0.045 eV in the
photon energy range 44.24 -- 69.74 eV.  High-resolution measurements near the ground-state ionization threshold were performed
at a photon energy resolution of 0.004 eV. The ground-state ionization threshold was determined
to be 45.717 $\pm$ 0.030 eV, which is 0.089 eV lower than the value tabulated in the NIST database \cite{NIST}.
A non-zero photoionization cross section below the $3s^23p^5(^2P^o_{1/2})$ metastable state was observed
that is attributed to ionization from higher-lying metastable states. The Cowan Hartree-Fock atomic
structure code was used to perform atomic-structure calculations to guide the assignments of the resonant features to Rydberg series.
Eight Rydberg series of $3p \rightarrow nd$ resonances originating from both the $3s^23p^5(^2P^o_{1/2})$
metastable state and $3s^23p^5(^2P^o_{3/2})$ ground state  were identified and
spectroscopically assigned using the quantum defect theory. Eight more Rydberg series of resonances due to $3s \rightarrow np$
inner-shell excitations were also identified. These series (due to  $3p \rightarrow nd$ and $3s \rightarrow np$ transitions) are tabulated according to
their measured energy positions, quantum defect parameters, series limits, and assignments. The limits for the assigned Rydberg series
provide additional evidence that the ground-state ionization potential of K$^{2+}$ is 45.717 eV.

Detailed calculations using the Breit-Pauli approximations within the R-matrix approach were performed
from the ground state and all metastable states lying below the $3s^23p^4(^3P)4s(^4P)$ levels over the
photon energy range 20 -- 70 eV. Suitable agreement with experiment is found with the intermediate coupling R-matrix
calculations using a non-statistical initial distribution among the metastable and ground states of the system.
The present semi-relativistic R-matrix calculations are consistent with 25\% of the parent K$^{2+}$ ion beam in the ground state,
25\% in the  $3s^23p^5(^2P^o_{1/2})$ metastable state the remaining 50\% distributed among the high-lying
metastable states considered for this system up to the $3s^23p^4(^3P)4s(^4P)$ levels.

The photoionization cross-sections from the present study are suitable for inclusion into
state-of-the-art photoionization modeling codes such as CLOUDY  \cite{ferland1998,ferland2003},
XSTAR \cite{kallman2001} and AtomDB \cite{Foster2012}  that are used to numerically
simulate the thermal and ionization structure of ionized astrophysical nebulae.

%
%
%

\begin{acknowledgments}
The Division of Chemical Sciences, Geosciences, and Biosciences of the U.S. Department of Energy supported this
research under grant DE-FG02-03ER15424 and contract DE-AC03-76SF-00098.  C. C. acknowledges
support from PAPIT-UNAM  IN107912-IN10261, Mexico. B. M. McL. acknowledges support by the U.S. National Science
Foundation, under the visitors program,  through a grant to ITAMP at the Harvard-Smithsonian Center for Astrophysics
where this work was completed and a visiting research fellowship from Queen's University Belfast.
The computational work was performed at the National Energy Research Scientific Computing Center in Oakland,
CA, and on the Kraken XT5 facility at the National Institute for Computational Science (NICS) in Knoxville,
TN.  The Kraken XT5 facility is a resource of the Extreme Science and Engineering Discovery Environment
(XSEDE), which is supported by National Science Foundation grant number OCI-1053575. The Advanced Light Source,
is supported by the Director, Office of Science, Office of Basic Energy Sciences of the US Department of Energy under
Contract DE-AC02-05CH11231.
\end{acknowledgments}
%
%
%
%
\bibliographystyle{apsrev4-1}
\bibliography{k2plus}
%
\section*{Appendices}

\renewcommand{\thefigure}{\Alph{subsection}.\arabic{figure}}
\renewcommand{\thetable}{\Alph{subsection}.\arabic{table}}
\setcounter{figure}{0}
\setcounter{table}{0}
\subsection{Quantum-defect analyses of Rydberg series}
%
%
%
\begin{table*}
\caption{\label{tab1} Principal quantum numbers $n$, resonance energies $E_n$ (eV), series limit $E_{\infty}$ (eV),  and
quantum defects $\delta_n$ of the K$^{2+}[3s^23p^4 (^1D_2)]nd (^2P,~^2D)$ series estimated from the experimental measurements.
Resonance energies are calibrated to within $\pm$0.030 eV and the mean quantum defects $\delta$ have an estimated uncertainty of $\pm$10\%. }
\begin{ruledtabular}
\begin{tabular}{ccccc}
               &          & Rydberg series 		&          		 			& Rydberg series 	   	 \\
               &          &   $^1$D$_2$    		&          		 			&     	         				\\
 Initial state &  $n$     & $E_{n}$ (eV)  	 &  $\delta_n$   	 			&      					\\
\hline                                                                            										 \\
                                                                                  							    		              \\
 $^2$P$^{\circ}_{1/2}$ &    			&        		 			&          			     \\
               &  [9$d$]       &  45.800           	&   0.478 $\pm$ 0.064   		 &        			     \\
               &  10             &   46.118       		&   0.539 $\pm$ 0.064		 & $[3s^23p^4 (^1D_2)]nd (^2P)$ \\
               &  11        	&  46.381       		&   0.475  $\pm$ 0.064		 &           			     \\
               &  12         	&  46.567        		&   0.454 $\pm$ 0.064 		 &        			    \\
               & $\cdot$  	& $\cdot$        		& $\cdot$ 		 			&				    \\
            & $\infty$  & 47.486 $\pm$ 0.038     &   $\delta$ = 0.527 $\pm$ 0.064      		 			&          		    	    \\

 $^2$P$^{\circ}_{1/2}$ &                		&          					 &                		         	 \\
               &  [9$d$]        &  45.805          	&  0.557 $\pm$ 0.062  		 &                		        	 \\
               &  10               &  46.133          	&  0.612  $\pm$ 0.062   		 &                		       	 \\
               &  11              &   46.406         	&  0.528  $\pm$ 0.062   		 & $[3s^23p^4 (^1D_2)]nd (^2D)$ \\
               &  12              &  46.588		         	&  0.555  $\pm$ 0.062   		 &                		    \\
               & $\cdot$      & $\cdot$        		& $\cdot$ 		 			&                			    \\
               & $\infty$      & 47.522 $\pm$ 0.038        & $\delta$ = 0.601 $\pm$ 0.062         		 		&               		 	    \\
               \\
 $^2$P$^{\circ}_{3/2}$ &    			&        		 			&          		 	     \\
              &  [9$d$]       	&46.066        		& 0.444  $\pm$ 0.037 		&          			     \\
               & 10      	 &46.375          		& 0.523  $\pm$ 0.037 		&           			     \\
               & 11       	& 46.633         		& 0.478  $\pm$ 0.037 		&         			     \\
               & 12       	&  46.810      		& 0.520  $\pm$ 0.037 		&  $[3s^23p^4 (^1D_2)]nd (^2P)$ \\
               & 13       	& 46.966   		& 0.411  $\pm$ 0.037 		&       			   \\
               & 14       	& 47.077             	& 0.395  $\pm$ 0.037 		&       			   \\
               & 15       	& 47.163            	& 0.415  $\pm$ 0.037   		&        			   \\
               & $\cdot$     & $\cdot$        		& $\cdot$  				& 				  \\
               & $\infty$ & 47.739 $\pm$ 0.034  &  $\delta$ =  0.506 $\pm$ 0.037       					&          			   \\
  $^2$P$^{\circ}_{3/2}$ &    			&        		 			&          		 	     \\
              &  [9$d$]       & 46.073        		& 0.451 $\pm$ 0.012   		&          			     \\
               & 10      	& 46.401       		& 0.466  $\pm$ 0.012    		&          			    \\
               & 11       	&  46.648       		& 0.448  $\pm$ 0.012  		&          			     \\
               & 12       	&  46.830     		& 0.451  $\pm$ 0.012 	 	& $[3s^23p^4 (^1D_2)]nd (^2D)$  \\
               & 13       	&  46.977     		& 0.401 $\pm$ 0.012  		&       			    \\
               & 14       	&  47.083           	& 0.435  $\pm$ 0.012       		&       			    \\
               & 15       	&  47.168            	& 0.465 $\pm$ 0.012         		&       			    \\
               & $\cdot$     & $\cdot$        		& $\cdot$  				&				    \\
               & $\infty$ & 47.748 $\pm$ 0.034  &    $\delta$ = 0.488 $\pm$ 0.012     					&  		       	 	     \\
                \\
\end{tabular}
\end{ruledtabular}
\end{table*}
%
%
\begin{table*}
\caption{\label{tab2} Principal quantum numbers $n$, resonance energies $E_n$ (eV), series limit $E_{\infty}$ (eV),  and
quantum defects $\delta_n$ of the K$^{2+}[3s^23p^4 (^1S_o)]nd (^2D_{3/2,5/2})$ series estimated from the experimental measurements.
Resonance energies are calibrated to $\pm$0.030 eV and the mean quantum defects $\delta$ are estimated with an uncertainty of $\pm$10\%. }
\begin{ruledtabular}
\begin{tabular}{ccccc}
               &          & Rydberg series 		&          		 			& Rydberg series 	   	 \\
               &          &   $^1$S$_0$    		&          		 			&     	         				\\
 Initial state &  $n$     & $E_{n}$ (eV)  	 &  $\delta_n$   	 			&      					\\
\hline                                                                            										 \\
 $^2$P$^{\circ}_{1/2}$ &                		&          					 &                		         	 \\
               &  [6$d$]        &  46.229          	&  0.481 $\pm$ 0.014  		 &                		        	 \\
               &  7                &  47.396        	 	&  0.450  $\pm$ 0.014   		 &                		       	 \\
               &  8                &   48.093         		&  0.466  $\pm$ 0.014   		 & $[3s^23p^4 (^1S_0)]nd (^2D_{3/2})$ \\
               &  9                &  48.562		         	&  0.481  $\pm$ 0.014   		 &                		    \\
               &  10              &  48.901		         	&  0.473  $\pm$ 0.014   		 &                		    \\
               & $\cdot$      & $\cdot$        		& $\cdot$ 		 			&                			    \\
               & $\infty$      & 50.249 $\pm$ 0.036        & $\delta$ = 0.496 $\pm$ 0.014         		 		&               		 	    \\
                                                                                  							    		              \\
 $^2$P$^{\circ}_{1/2}$ &    			&        		 			&          			     \\
               &  [6$d$]        &  46.188          	&  0.481 $\pm$ 0.009  		 &                		        	 \\
               &  7                &  47.335        	 	&  0.472  $\pm$ 0.009   		 &                		       	 \\
               &  8                &   48.037         		&  0.491  $\pm$ 0.009   		 & $[3s^23p^4 (^1S_0)]nd (^2D_{5/2})$ \\
               &  9                &  48.522		         	&  0.480  $\pm$ 0.009   		 &                		    \\
               &  10              &  48.870		         	&  0.436  $\pm$ 0.009   		 &                		    \\
               & $\cdot$      & $\cdot$        		& $\cdot$ 		 			&                			    \\
           & $\infty$  & 50.209 $\pm$ 0.034     &   $\delta$ = 0.502 $\pm$ 0.009      		 			&          		    	    \\

  $^2$P$^{\circ}_{3/2}$ &    			&        		 			&          		 	     \\
              &  [6$d$]       	 & 46.507        		& 0.469 $\pm$ 0.012   		&          			     \\
               & 7      	 	&  47.663       		& 0.440  $\pm$ 0.012    		&          			    \\
               & 8      	 	&  48.355       		& 0.460  $\pm$ 0.012  		&  $[3s^23p^4 (^1S_0)]nd (^2D_{3/2})$        			     \\
               & 9      	 	&  48.830    		& 0.460  $\pm$ 0.012 	 	&  				 \\
               & 10       	&  49.163     		& 0.461 $\pm$ 0.012  		&       			    \\
               & $\cdot$     & $\cdot$        		& $\cdot$  				&				    \\
               & $\infty$ & 50.509 $\pm$ 0.035  &    $\delta$ = 0.484 $\pm$ 0.012					&  		       	 	     \\
  $^2$P$^{\circ}_{3/2}$ &    			&        		 			&          		 	     \\
              &  [6$d$]       &46.431        		& 0.504  $\pm$ 0.009 		&          			     \\
               & 7      	 	&47.603          		& 0.482  $\pm$ 0.009 		&           			     \\
               & 8       		& 48.310         		& 0.498  $\pm$ 0.009 		& $[3s^23p^4 (^1S_0)]nd (^2D_{5/2})$         			     \\
               & 9       		& 48.789      		& 0.502  $\pm$ 0.009 		&  			\\
               & 10       	& 49.133   		& 0.483  $\pm$ 0.009 		&       			   \\
                & $\cdot$     & $\cdot$        		& $\cdot$  				& 				  \\
               & $\infty$ &50.485 $\pm$ 0.034  &  $\delta$ =  0.521 $\pm$ 0.009       					&          			   \\
                              \\
\end{tabular}
\end{ruledtabular}
\end{table*}
\begin{table*}
\caption{\label{tab3} Principal quantum numbers $n$, resonance energies $E_n$ (eV), series limit $E_{\infty}$ (eV),  and
			quantum defects $\delta_n$ of the K$^{2+}[3s^23p^5(^2P^o_{1/2})  \rightarrow 3s3p^5 (^3P^o~_{2,1,0})np]$
			series estimated from the experimental measurements. Resonance energies are calibrated to $\pm$0.030 eV
			and the mean quantum defects $\delta$ are estimated with an uncertainty of $\pm$10\%. }
\begin{ruledtabular}
\begin{tabular}{ccccc}
               &          & Rydberg series 		&          		 			& Rydberg series 	   	 \\
               &          &   $^3$P$_{2,1,0}$    		&          		 			&     	         				\\
 Initial state &  $n$     & $E_{n}$ (eV)  	 &  $\delta_n$   	 			&      					\\
\hline                                                                            										 \\
 $^2$P$^{\circ}_{1/2}$ &                		&          					 &                		         	 \\
               &  [4$p$]        &  53.938          	&  0.171 $\pm$  0.002  		 &                		        	 \\
               &  5                 &  57.061        	 	&  0.161  $\pm$ 0.002   		 &                		       	 \\
               &  6                 &  58.701          	&  0.160  $\pm$ 0.002   		 & $[3s3p^5 (^3P^o_2)]np$ \\
               &  7                 &  59.671	         	&  0.165  $\pm$ 0.002   		 &                		    \\
               &  8	           &  60.300	         	&  0.158  $\pm$ 0.002   		 &                		    \\
               &  9	           &  60.715	         	&  0.186  $\pm$ 0.002   		 &                		    \\
               &  10	           &  61.035	         	&  0.127  $\pm$ 0.002   		 &                		    \\
               &  11	           &  61.255	         	&  0.130  $\pm$ 0.002   		 &                		    \\
               & $\cdot$       & $\cdot$        		& $\cdot$ 		 			&                			    \\
               & $\infty$       & 62.291 $\pm$ 0.034        & $\delta$ = 0.185 $\pm$ 0.002         		 		&               		 	    \\
                                                                                  							    		              \\
$^2$P$^{\circ}_{1/2}$ &                		&          					 &                		         	 \\
               &  [4$p$]        &  54.032          	&  0.171 $\pm$  0.004  		 &                		        	 \\
               &  5                 &  57.161        	 	&  0.161  $\pm$ 0.004   		 &                		       	 \\
               &  6                 &  58.806          	&  0.160  $\pm$ 0.004   		 & $[3s3p^5 (^3P^o_1)]np$ \\
               &  7                 &  59.770	         	&  0.165  $\pm$ 0.004   		 &                		    \\
               &  8	           &  60.390	         	&  0.158  $\pm$ 0.004   		 &                		    \\
               &  9	           &  60.805	         	&  0.186  $\pm$ 0.004   		 &                		    \\
               &  10	           &  61.105	         	&  0.127  $\pm$ 0.004   		 &                		    \\
               &  11	           &  61.335	         	&  0.130  $\pm$ 0.004   		 &                		    \\
               & $\cdot$       & $\cdot$        		& $\cdot$ 		 			&                			    \\
               & $\infty$       & 62.378 $\pm$ 0.034        & $\delta$ = 0.181 $\pm$ 0.004         		 		&               		 	    \\
                                                                                  							    		              \\
$^2$P$^{\circ}_{1/2}$ &                		&          					 &                		         	 \\
               &  [4$p$]        &  54.472          	&  0.143 $\pm$  0.023  		 &                		        	 \\
               &  5                 &  57.431        	 	&  0.181  $\pm$ 0.023   		 &                		       	 \\
               &  6                 &  59.166          	&  0.118  $\pm$ 0.023   		 & $[3s3p^5 (^3P^o_0)]np$ \\
               &  7                 &  60.125	         	&  0.110  $\pm$ 0.023   		 &                		    \\
               & $\cdot$       & $\cdot$        		& $\cdot$ 		 			&                			    \\
               & $\infty$       & 62.705 $\pm$ 0.067        & $\delta$ = 0.164 $\pm$ 0.023         		 		&               		 	    \\
                                                                                  							    		              \\
\end{tabular}
\end{ruledtabular}
\end{table*}
\begin{table*}
\caption{\label{tab4} Principal quantum numbers $n$, resonance energies $E_n$ (eV), series limit $E_{\infty}$ (eV),  and
quantum defects $\delta_n$ of the K$^{2+}[3s^23p^5(^2P^o_{3/2})  \rightarrow 3s3p^5 (^3P^o~_{2,1,0})np]$
series estimated from the experimental measurements.
Resonance energies are calibrated to $\pm$0.030 eV and the mean quantum
defects $\delta$ are estimated to within an error of 10\%. }
\begin{ruledtabular}
\begin{tabular}{ccccc}
               &          & Rydberg series 		&          		 			& Rydberg series 	   	 \\
               &          &   $^3$P$_{2,1,0}$    		&          		 			&     	         				\\
 Initial state &  $n$     & $E_{n}$ (eV)  	 &  $\delta_n$   	 			&      					\\
\hline                                                                            										 \\
 $^2$P$^{\circ}_{3/2}$ &                		&          					 &                		         	 \\
               &  [4$p$]        &  54.162          	&  0.134 $\pm$  0.004  		 &                		        	 \\
               &  5                 &  57.161        	 	&  0.145  $\pm$ 0.004   		 &                		       	 \\
               &  6                 &  58.806          	&  0.127  $\pm$ 0.004   		 & $[3s3p^5 (^3P^o_2)]np$ \\
               &  7                 &  59.770	         	&  0.119  $\pm$ 0.004   		 &                		    \\
               &  8	           &  60.390	         	&  0.108  $\pm$ 0.004   		 &                		    \\
               &  9	           &  60.805	         	&  0.115  $\pm$ 0.004   		 &                		    \\
                & $\cdot$       & $\cdot$        		& $\cdot$ 		 			&                			    \\
               & $\infty$       & 62.356 $\pm$ 0.034        & $\delta$ = 0.151 $\pm$ 0.004         		 		&               		 	    \\
                                                                                  							    		              \\
$^2$P$^{\circ}_{3/2}$ &                		&          					 &                		         	 \\
               &  [4$p$]        &  54.242          	&  0.157 $\pm$  0.004  		 &                		        	 \\
               &  5                 &  57.291        	 	&  0.168  $\pm$ 0.004   		 &                		       	 \\
               &  6                 &  58.961          	&  0.147  $\pm$ 0.004   		 & $[3s3p^5 (^3P^o_1)]np$ \\
               &  7                 &  59.930	         	&  0.143  $\pm$ 0.004   		 &                		    \\
               &  8	           &  60.555	         	&  0.135  $\pm$ 0.004   		 &                		    \\
               &  9	           &  60.975	         	&  0.139  $\pm$ 0.004   		 &                		    \\
                & $\cdot$       & $\cdot$        		& $\cdot$ 		 			&                			    \\
               & $\infty$       & 62.535 $\pm$ 0.034        & $\delta$ = 0.174 $\pm$ 0.004         		 		&               		 	    \\
                                                                                  							    		              \\
$^2$P$^{\circ}_{3/2}$ &                		&          					 &                		         	 \\
               &  [4$p$]        &  54.202          	&  0.169 $\pm$  0.002  		 &                		        	 \\
               &  5                 &  57.301        	 	&  0.168  $\pm$ 0.002   		 &                		       	 \\
               &  6                 &  58.961          	&  0.156  $\pm$ 0.002   		 & $[3s3p^5 (^3P^o_0)]np$ \\
               &  7                 &  59.930	         	&  0.159  $\pm$ 0.002   		 &                		    \\
               &  8                 &  60.555	         	&  0.159  $\pm$ 0.002   		 &                		    \\
               & $\cdot$       & $\cdot$        		& $\cdot$ 		 			&                			    \\
               & $\infty$       & 62.547 $\pm$ 0.034        & $\delta$ = 0.184 $\pm$ 0.002         		 		&               		 	    \\
                                                                                  							    		              \\
\end{tabular}
\end{ruledtabular}
\end{table*}
\begin{table*}
\caption{\label{tab5} Principal quantum numbers $n$, resonance energies $E_n$ (eV), series limit $E_{\infty}$ (eV),  and
quantum defects $\delta_n$ of the K$^{2+}[3s3p^5 (^1P^o~_1)]np$ series estimated from the experimental measurements.
Resonance energies are calibrated to $\pm$0.030 eV and the mean quantum defects $\delta$ are estimated to within an error of 10\%. }
\begin{ruledtabular}
\begin{tabular}{ccccc}
               &          & Rydberg series 			&          		 			& Rydberg series 	   	 \\
               &          &   $^1$P$_1$    			&          		 			&     	         				\\
 Initial state &  $n$     & $E_{n}$ (eV)  	 	&  $\delta_n$   	 			&      					\\
\hline                                                                            										 \\
 $^2$P$^{\circ}_{1/2}$ 	&                			&          					 &                		         	 \\
               &  [4$p$]        	&  50.664          		&  1.262 $\pm$ 0.006  		 &                		        	 \\
               &  5               		&  58.241          		&  1.260  $\pm$ 0.006   		 &    $[3s3p^5 (^1P^o_1)]np$        \\
               &  6              		&  61.510         		&  1.275  $\pm$ 0.006   		 &  					 \\
               &  7              		&  63.334		         &  1.215  $\pm$ 0.006   		 &                		    \\
               & $\cdot$     		 & $\cdot$        		& $\cdot$ 		 			&                			    \\
               & $\infty$      	& 66.993 $\pm$ 0.049        & $\delta$ = 1.272 $\pm$ 0.006        &               		 	    \\
                                                                                  							    		              \\
  $^2$P$^{\circ}_{3/2}$ &    				&		        		 		&          		 	     \\
              &  [4$p$]       		& 50.949        		& 1.240 $\pm$ 0.006  		&          			     \\
               & 5      		 	& 58.396       		& 1.231  $\pm$ 0.006    		& $[3s3p^5 (^1P^o_1)]np$   \\
               & 6       			& 61.580      		& 1.254  $\pm$ 0.006  		&          			     \\
               &  7              		& 63.374		         &  1.202  $\pm$ 0.006   		&                		    \\
               & $\cdot$     		& $\cdot$        		& $\cdot$  				&				    \\
               & $\infty$ 		& 67.017 $\pm$ 0.049  &    $\delta$ = 1.250 $\pm$ 0.006     	&  		       	 	     \\
               \\
\end{tabular}
\end{ruledtabular}
\end{table*}

%
\begin{figure*}
\begin{center}
 \includegraphics[width=\textwidth]{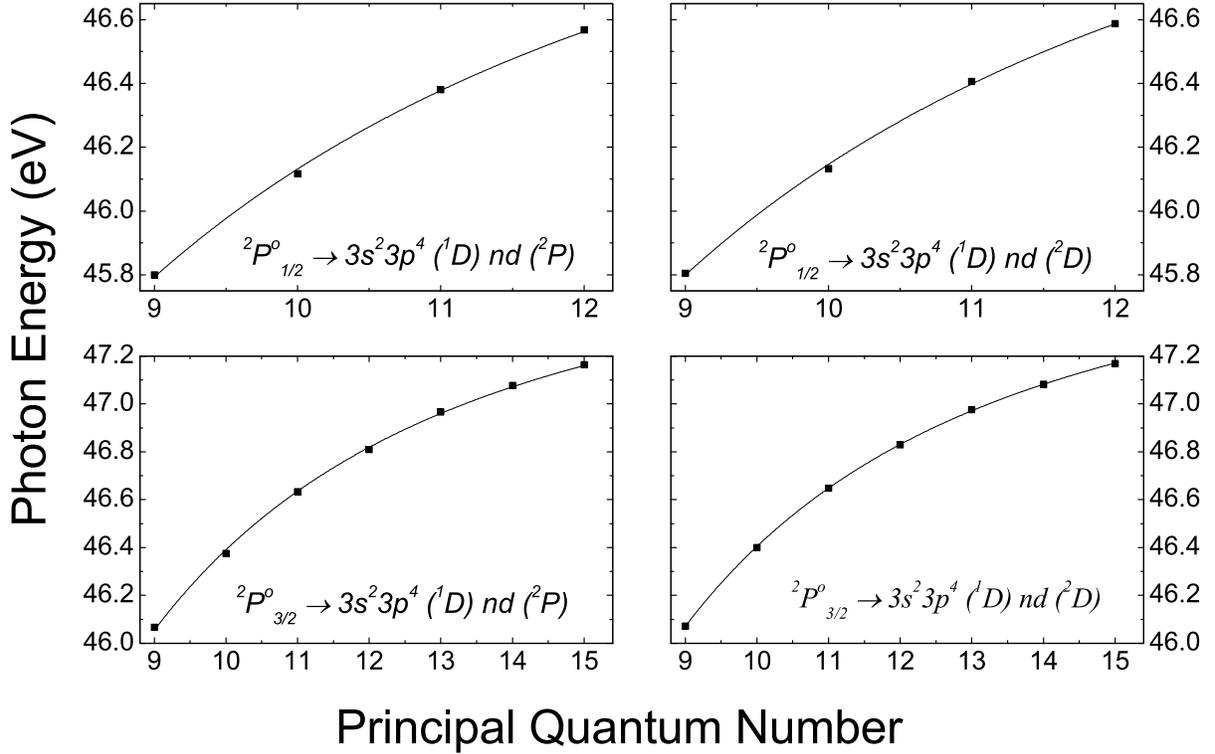}
 \caption{\label{rydberg1} Rydberg fits for the $3s^23p^4(^1D_2)nd$ series originating from both the $^2P^o_{3/2}$
					ground state and the $^2P^o_{1/2}$ metastable state.}
\end{center}
\end{figure*}
%
%
\begin{figure*}
\begin{center}
 \includegraphics[width=\textwidth]{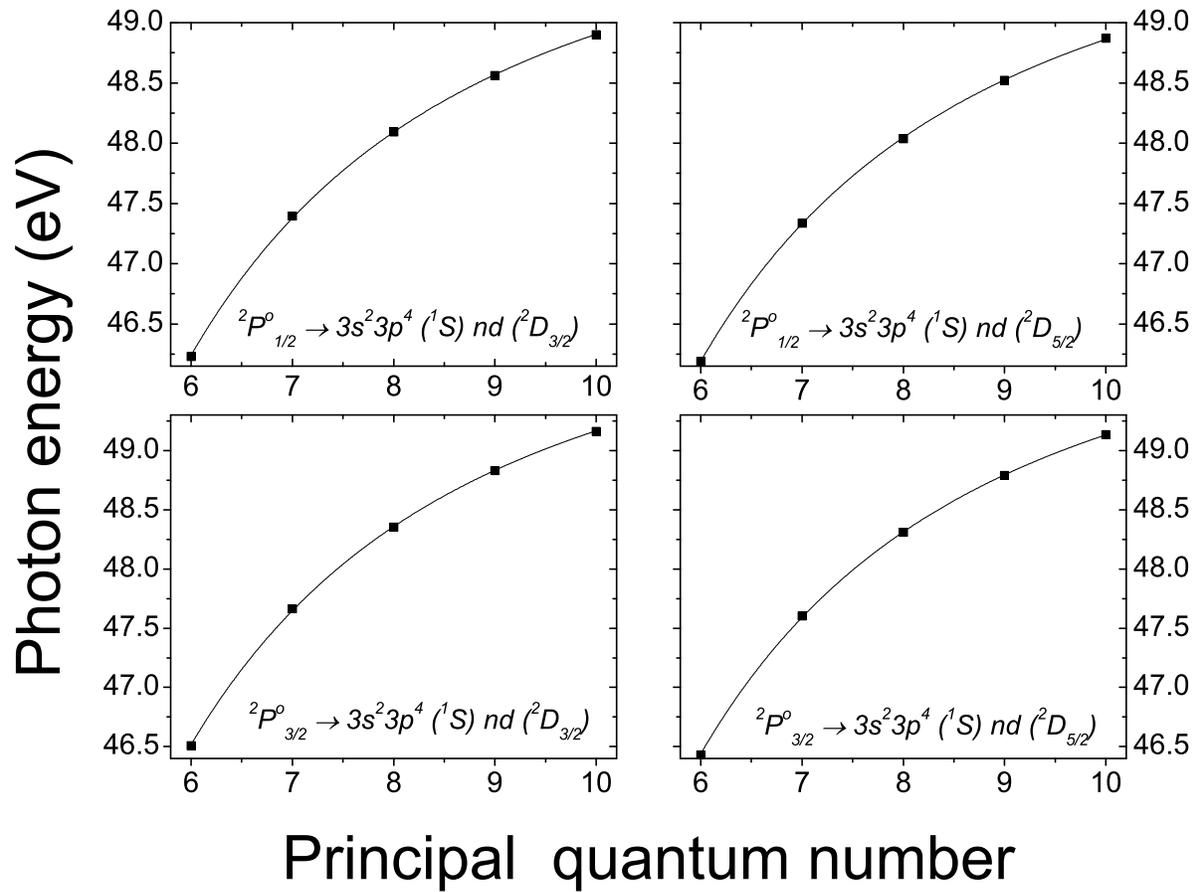}
 \caption{\label{rydberg2} Fits of the Rydberg formula for the $3s^23p^4(^1S_0)nd$ resonance energies for series
					originating from the $^2P^o_{3/2}$ ground state and the $^2P^o_{1/2}$ metastable state.}
\end{center}
\end{figure*}
%
%
\begin{figure*}
\begin{center}
 \includegraphics[width=\textwidth]{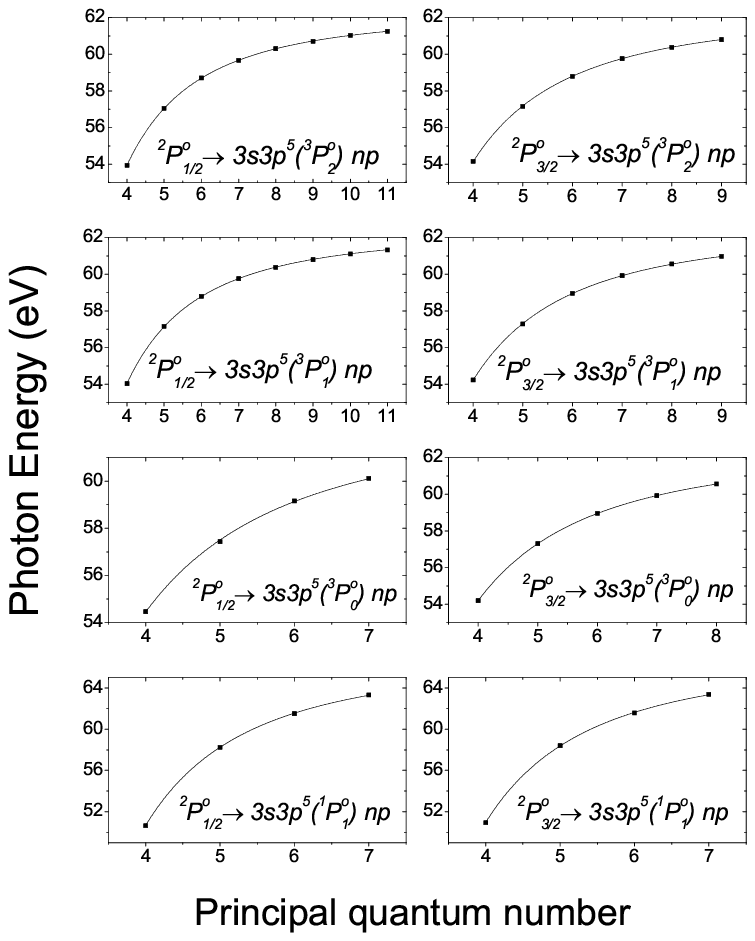}
 \caption{\label{8series2} Fits of Rydberg formula for the $3s3p^5(^3P^o~_{2,1,0})np$ and $3s3p^5(^1P^o~_1)np$ series
					originating from the $^2P^o_{3/2}$ ground state and the $^2P^o_{1/2}$ metastable state.}
\end{center}
\end{figure*}

%
%
%

\subsection{R-Matrix theory results for the ground and metastable initial states}
%
\setcounter{figure}{0}
%
%
\begin{figure*}
\begin{center}
\includegraphics[width=\textwidth, angle=-90,scale=0.7]{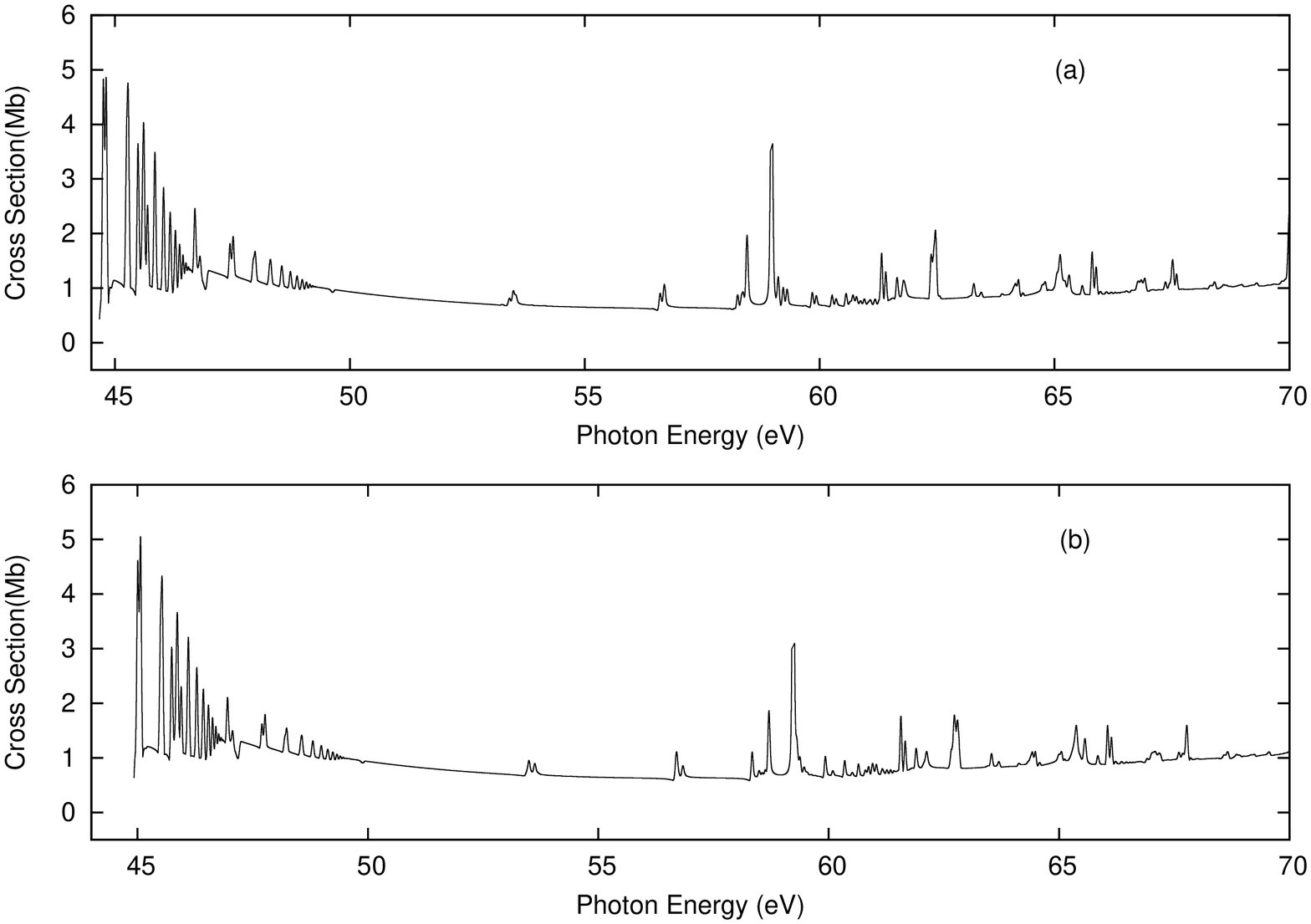}
\caption{\label{rmat11}   Breit-Pauli R-matrix photoionization cross section calculations
				  for the metastable  (a) $3s^23p^5~^2P^{\circ}_{1/2}$
				   and (b) $3s^23p^5~^2P^{\circ}_{3/2}$ ground state of K$^{2+}$.
				   The theoretical results has been convoluted with a Gaussian
				   of 45 meV FWHM to simulate the photon energy resolution of the experiment.}
\end{center}
\end{figure*}
%

%
\begin{figure*}
\begin{center}
\includegraphics[width=\textwidth, angle=-90,scale=0.7]{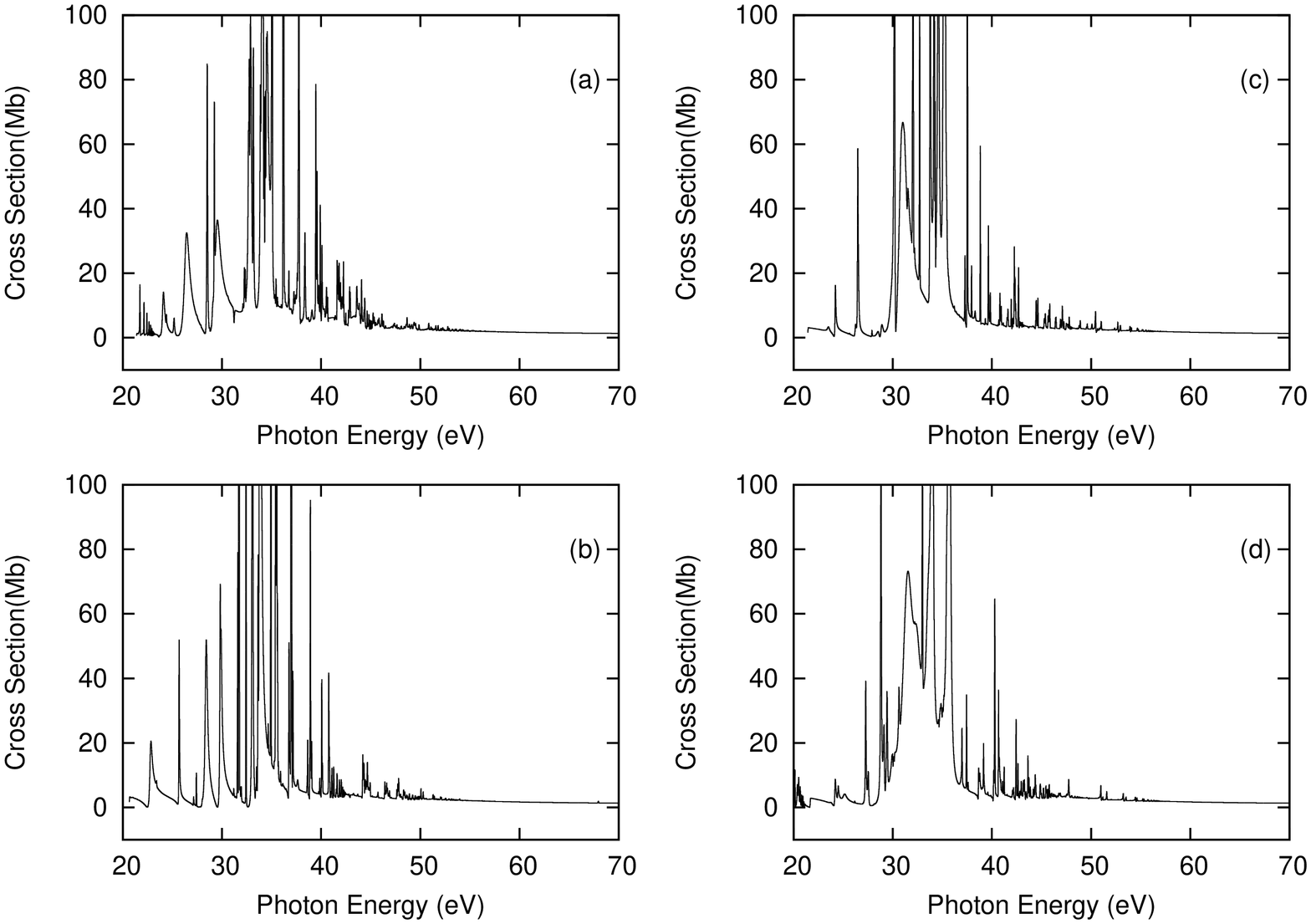}
\caption{\label{rmat12}   Breit-Pauli R-matrix photoionization  cross section calculations
					for the (a) $3d ~^4P_{1/2}$ and (b) $4s~^4P_{1/2}$, (c) $3d~^4F_{9/2}$  			
				     and (d) $3d~^2G_{9/2}$  metastable states of K$^{2+}$ as listed in Table \ref{dipole}.
				     The theoretical results has been convoluted with a Gaussian
				     of 45 meV FWHM to simulate the photon energy resolution of the experiment.}
\end{center}
\end{figure*}
%
%
\begin{figure*}
\begin{center}
\includegraphics[width=\textwidth,angle=-90,scale=0.7]{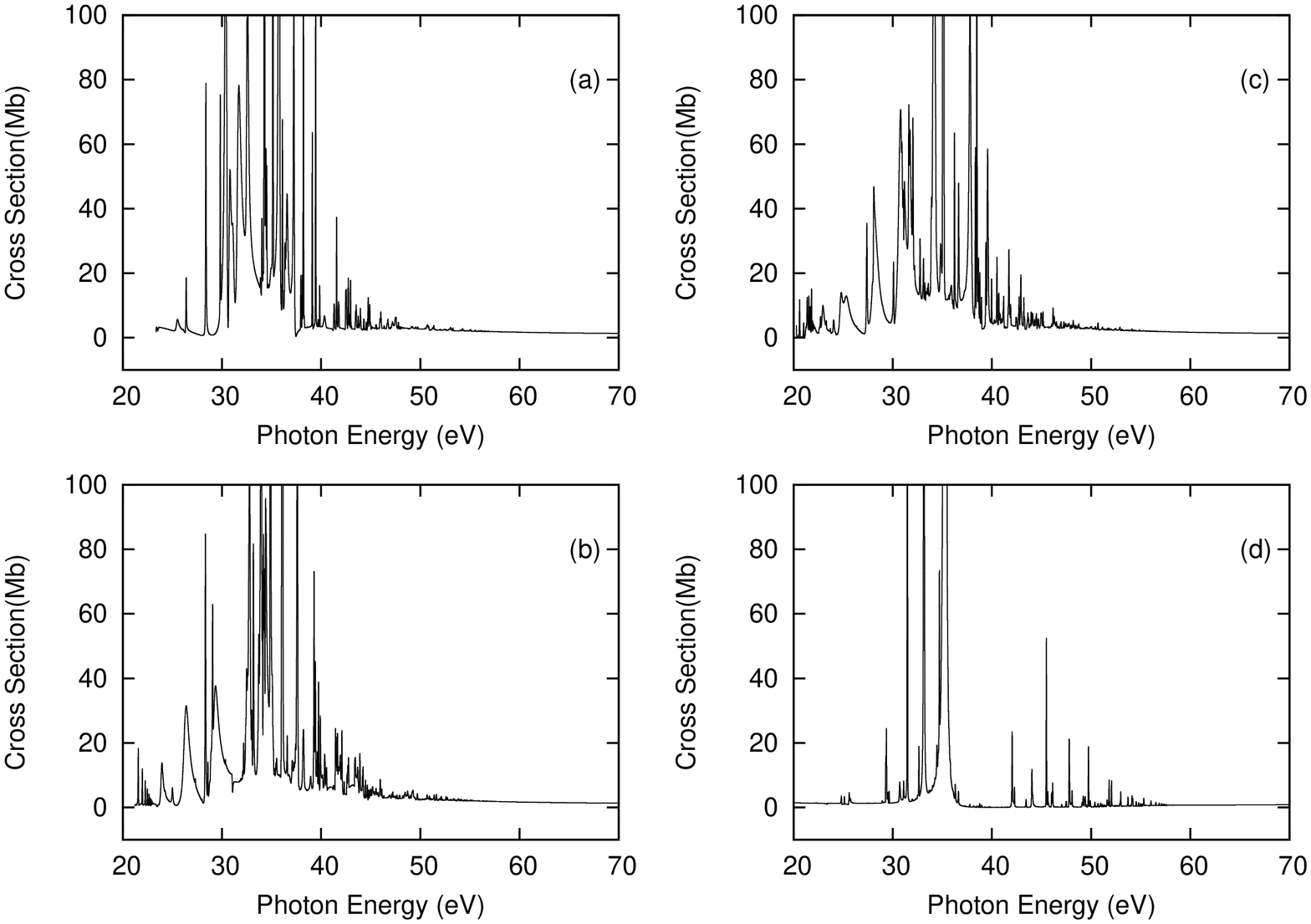}
\caption{\label{rmat13}   Breit-Pauli R-matrix photoionization  cross section calculations for the (a) $3d ~^4D_{3/2}$, (b) $3d~^4P_{3/2}$  			
				     (c) $4s~^4P_{3/2}$ and (d) $4s~^4P_{5/2}$  metastable states of K$^{2+}$ as listed in Table \ref{dipole}.
				     The theoretical results has been convoluted with a Gaussian
				     of 45 meV FWHM to simulate the photon energy resolution of the experiment.}
\end{center}
\end{figure*}
%
 %
%
\begin{figure*}
\begin{center}
\includegraphics[width=\textwidth,angle=-90,scale=0.7]{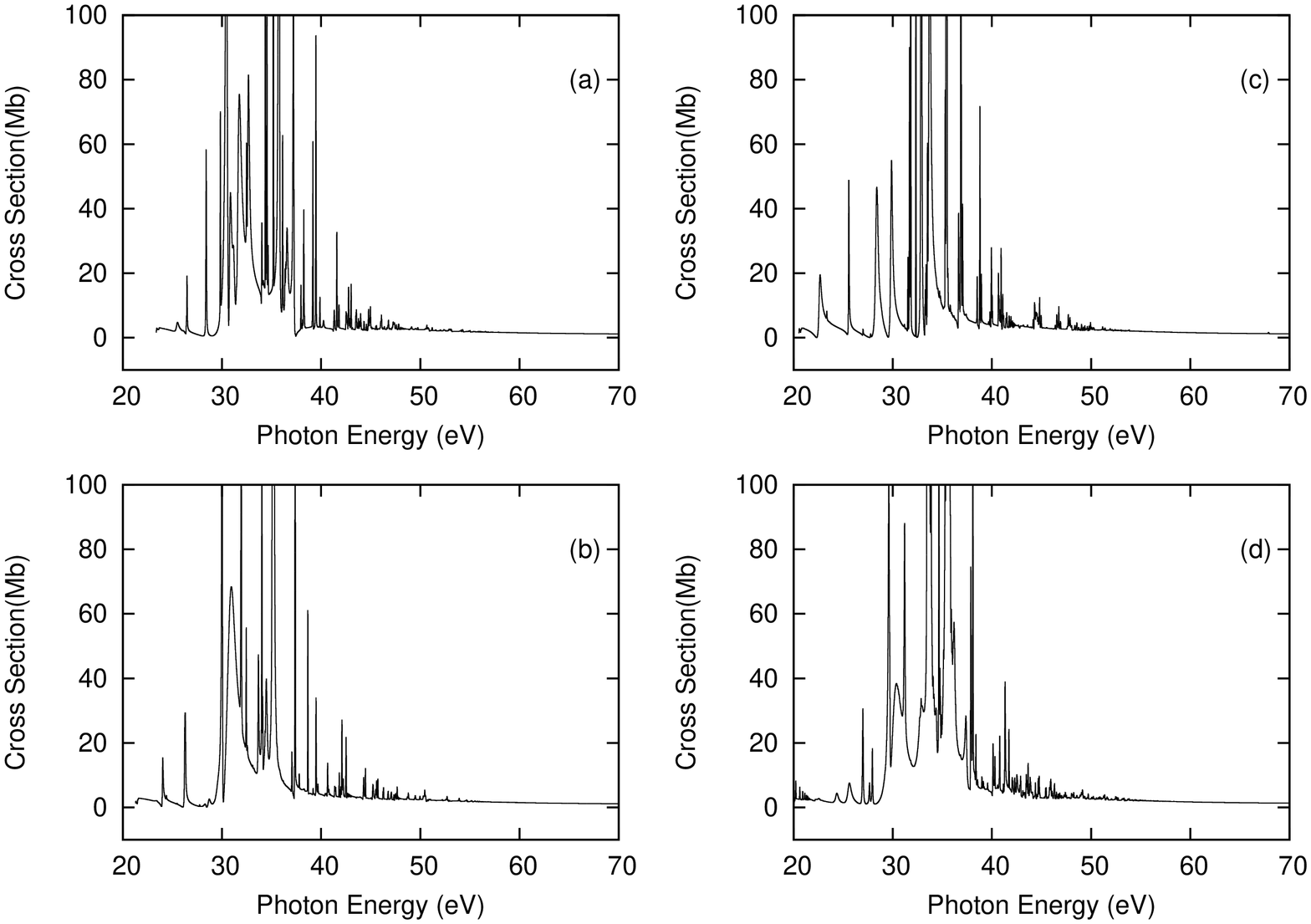}
\caption{\label{rmat14}   Breit-Pauli R-matrix photoionization  cross section calculations
 				     for the (a) $3d ~^4D_{5/2}$,  (b) $3d~^4F_{5/2}$,  	
				     (c) $3d~^4P_{5/2}$  and (d) $3d~^2F_{5/2}$  metastable states of K$^{2+}$ as listed in Table \ref{dipole}.
				     The theoretical results has been convoluted with a Gaussian
				     of 45 meV FWHM to simulate the photon energy resolution of the experiment.}
\end{center}
\end{figure*}
 %
%
\begin{figure*}
\begin{center}
\includegraphics[width=\textwidth,angle=-90,scale=0.7]{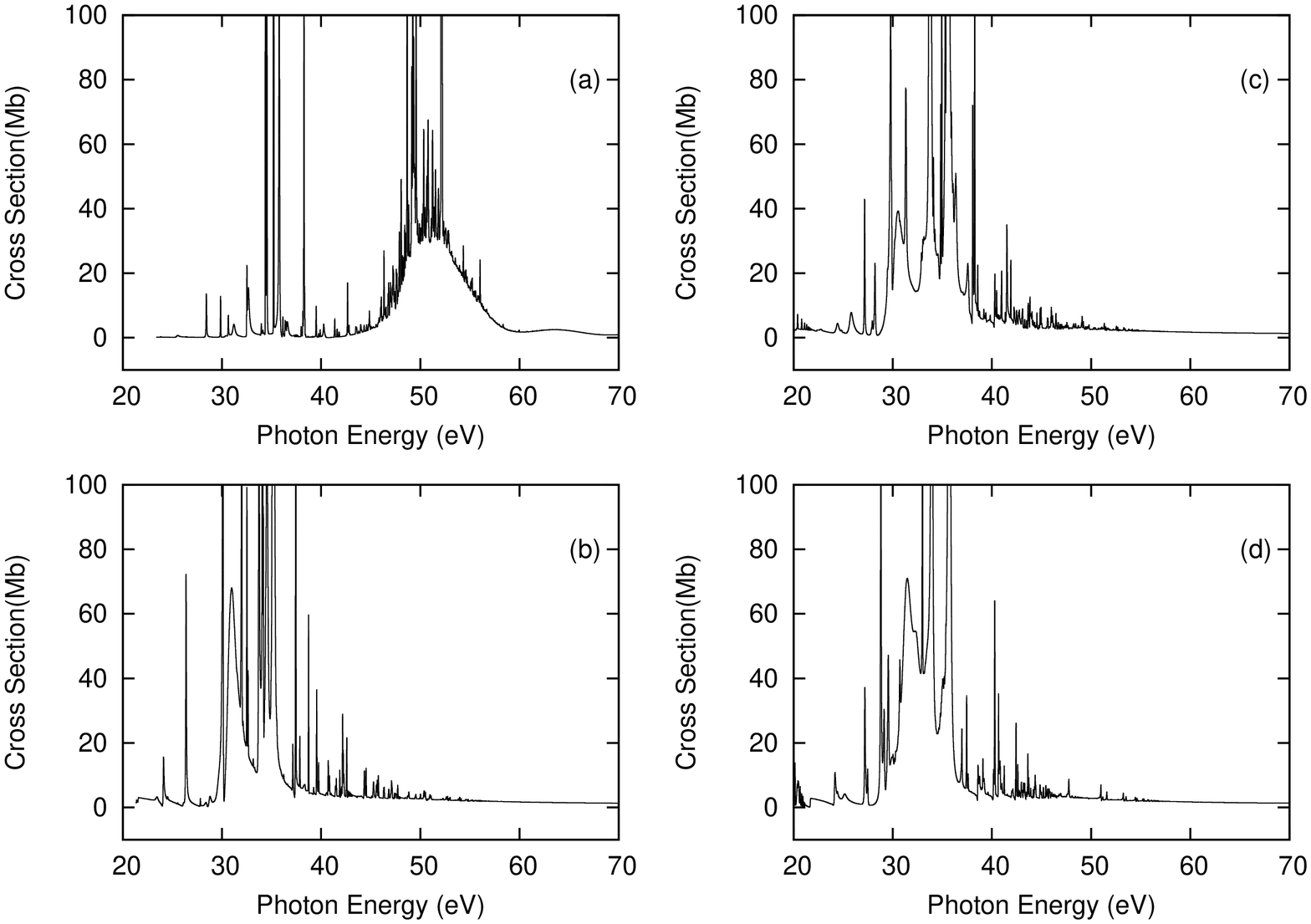}
\caption{\label{rmat15}   Breit-Pauli R-matrix photoionization  cross section calculations
				      for the (a) $3d ~^4D_{7/2}$, (b) $3d~^4F_{7/2}$,  	
				     (c) $3d~^2F_{7/2}$ and (d) $3d~^2G_{7/2}$  metastable states of K$^{2+}$ as listed in Table \ref{dipole}.
				     The theoretical results has been convoluted with a Gaussian
				     of 45 meV FWHM to simulate the photon energy resolution of the experiment.}
\end{center}
\end{figure*}
%
%
\end{document}